\documentclass[12pt,twoside, a4paper]{article}
\def\pd{\partial}
\def\mc{\mathcal}
\def\ul{\underline}

\usepackage[dvips]{graphicx}
\usepackage{amssymb}
\usepackage[bbgreekl]{mathbbol}
\usepackage{amssymb,amsmath}
\usepackage{graphicx}
\usepackage{dsfont}
\usepackage{caption}
\usepackage{subcaption}
\usepackage{mathtools}
\usepackage{verbatim}
\usepackage{graphicx}
\usepackage{multirow}
\usepackage[outline]{contour}
\usepackage{xcolor,colortbl}
\input{epsf.sty}  
\begin{document}
\begin{center}
\Large{\textbf{Holographic RG flows and Janus interfaces from ISO(3)$\times$U(1) $F(4)$ gauged supergravity}}
\end{center}
\vspace{1 cm}
\begin{center}
\large{\textbf{Parinya Karndumri}}
\end{center}
\begin{center}
String Theory and Supergravity Group, Department
of Physics, Faculty of Science, Chulalongkorn University, 254 Phayathai Road, Pathumwan, Bangkok 10330, Thailand \\
E-mail: parinya.ka@hotmail.com \vspace{1 cm}
\end{center}
\begin{abstract}
We study six-dimensional $F(4)$ gauged supergravity coupled to four vector multiplets with a non-semisimple $ISO(3)\times U(1)$ gauge group. This gauged supergravity arises from a consistent truncation of type IIB string theory on $S^2\times \Sigma$ with $\Sigma$ being a Riemann surface. We find that the gauged supergravity admits two $SO(3)$ symmetric $AdS_6$ vacua, one with $N=(1,1)$ supersymmetry and the other one non-supersymmetric. We compute all scalar masses at these critical points and find that only the supersymmetric $AdS_6$ vacuum is stable. By truncating to $SO(3)\times U(1)$ and $SO(2)\times U(1)$ invariant sectors, we study holographic RG flow solutions from this critical point to a number of non-conformal phases of the dual $N=2$ SCFT in five dimensions. Finally, we give some examples of supersymmetric Janus solutions describing conformal interfaces within the five-dimensional SCFT. All of these solutions provide first holographic results from matter-coupled $F(4)$ gauged supergravity with non-semisimple gauge groups.     
\end{abstract}
\newpage
\section{Introduction}
There is a unique, up to an additional flavor symmetry, $N=2$ superconformal symmetry in five dimensions consisting of $16$ supercharges \cite{Nahm_res}. The realization of this as conformal fixed points in strongly coupled five-dimensional field theories has been found long ago in \cite{Seiberg_5Dfield,Seiberg_5Dfield3,Seiberg_5Dfield2}. After the proposal of the AdS/CFT correspondence \cite{maldacena,Gubser_AdS_CFT,Witten_AdS_CFT}, these conformal fixed points have been interpreted in terms of the dual $AdS_6$ geometry in \cite{ferrara_AdS6}, and an $AdS_6\times S^4$ background arising from the near horizon geometry of D4-D8 brane system has been found in \cite{D4D8}. Recently, the study of $N=2$ five-dimensional superconformal field theories (SCFTs) has attracted a lot of attention leading to many new interesting results, see for example \cite{5DSymmetry_enhanced,Bergman,Bergman2,Bergman3,AdS6_CFT5_1,AdS6_CFT5_2,
AdS6_CFT5_3,AdS6_CFT5_4,AdS6_CFT5_5}. 
\\
\indent In this paper, we will study AdS$_6$/CFT$_5$ correspondence using the half-maximal $N=(1,1)$ gauged supergravity in six dimensions. Working in lower-dimensional gauged supergravity in contrast to string/M-theory in ten or eleven dimensions makes various holographic computations more traceable. The $N=(1,1)$ gauged supergravity is the only gauged supergravity in six dimensions that admits supersymmetric $AdS_6$ vacua in accordance with the fact that there exists only $N=2$ superconformal symmetry in five dimensions. The first $N=(1,1)$, or $F(4)$, gauged supergravity has been constructed long ago in \cite{F4_Romans}, and the generalization to matter-coupled theories has been constructed in \cite{F4SUGRA1} and \cite{F4SUGRA2}. 
\\
\indent The global symmetry of the $N=(1,1)$ supergravity coupled to $n$ vector multiplets is given by $\mathbb{R}^+\times SO(4,n)$. An example of compact $SU(2)\times G_c$ gauge group, with $SU(2)$ being a diagonal subgroup of $SU(2)\times SU(2)\sim SO(4)$ and a compact group $G_c\subset SO(n)$, has also been explicitly constructed in \cite{F4SUGRA1,F4SUGRA2}. Various results on holographic studies in this framework have appeared in a number of previous works, see for example \cite{F4_nunezAdS6,F4_flow,5DSYM_from_F4,6D_twist,6D_Janus,6D_Janus_RG,AdS6_BH_Minwoo1,
AdS6_BH_Zaffaroni,AdS6_BH_Minwoo,AdS6_BH,AdS6_defect}. All of these results have been found within six-dimensional gauged supergravities with compact gauge groups, $SU(2)$ for pure $F(4)$ and $SU(2)\times SU(2)$ for matter-coupled $F(4)$ gauged supergravities. The former is known to arise from a consistent truncation of massive type IIA theory on $S^4$ \cite{Massive_IIA_onS4}. On the other hand, the latter currently has no known embedding in string/M-theory. This renders the results from the $SU(2)\times SU(2)$ gauged supergravity of limited use in the holographic context. 
\\
\indent In the present paper, we will consider another matter-coupled $F(4)$ gauged supergravity with a non-semisimple $ISO(3)\times U(1)$ gauge group. This is obtained by coupling the $N=(1,1)$ supergravity multiplet to four vector multiplets leading to $\mathbb{R}^+\times SO(4,4)$ global symmetry. The $ISO(3)$ gauge group is embedded in $SO(4,3)\subset SO(4,4)$ with the remaining $U(1)$ factor corresponding to the $U(1)$ abelian gauge symmetry associated with the fourth vector multiplet of the ungauged theory. Accordingly, if we truncate out the fourth vector multiplet, which is always a consistent truncation, we obtain $ISO(3)$ $F(4)$ gauged supergravity coupled to three vector multiplets. According to the general results of \cite{AdS6_Jan}, both of these gauged supergravities admit supersymmetric $AdS_6$ vacua. Therefore, we will consider both of these cases in this paper. In addition, these gauged supergravities have been shown in \cite{Henning_Malek_AdS7_6} to arise from consistent truncations of type IIB theory on $S^2\times \Sigma$ with $\Sigma$ being a Riemann surface, and the corresponding $AdS_6$ vacua are identified with those found in \cite{AdS6_IIB1,AdS6_IIB2,AdS6_IIB3}.   
\\
\indent We will first study supersymmetric $AdS_6$ vacua of the $ISO(3)\times U(1)$ gauged supergravity. As we will see, it turns out that there is only one supersymmetric $AdS_6$ vacuum. This is expected to be dual to an $N=2$ SCFT in five dimensions. We compute the full scalar masses and find holographic RG flow solutions interpolating between this $AdS_6$ critical point and singular geometries corresponding to different non-conformal phases of the dual $N=2$ SCFT. Unlike the previously studied compact gauge groups, in which only relevant operators appear at the $AdS_6$ fixed point with all scalars vanishing, the scalar masses at the $AdS_6$ vacuum of the $ISO(3)\times U(1)$ gauged supergravity contain both relevant and irrelevant dual operators. 
\\
\indent We also consider supersymmetric Janus solutions interpolating between $AdS_6$ vacua at two limits. These solutions holographically describe half-supersymmetric conformal interfaces within the five-dimensional $N=2$ SCFT. Many aspects of the results are similar to those obtained from compact gauge groups although some features are distinct due to the different nature of irrelevant and relevant operators. This study should provide first holographic results on matter-coupled $F(4)$ gauged supergravity with non-semisimple gauge groups in contrast to all the previous results that are obtained exclusively from compact gaugings. 
\\
\indent The paper is organized as follows. We give a review of the general structure of $F(4)$ gauged supergravity coupled to vector multiplets for an arbitrary gauge group compatible with supersymmetry in section \ref{6D_SO4gaugedN2}. We then consider a non-semisimple gauge group $ISO(3)\times U(1)$ and look for possible supersymmetric $AdS_6$ vacua in section \ref{Vacua}. In section \ref{RG_flows}, holographic RG flow solutions are obtained using standard domain wall ansatz within the truncations to $SO(3)\times U(1)$ and $SO(2)\times U(1)$ invariant scalars. We then derive a relevant set of BPS equations for Janus solutions, and various examples of numerical regular Janus solutions are given in section \ref{Janus}. We end the paper by giving some conclusions and comments on the results in section \ref{conclusion}.

\section{Matter-coupled $F(4)$ gauged supergravity in six dimensions}\label{6D_SO4gaugedN2}
In this section, we give a review of matter-coupled $F(4)$ gauged supergravity in six dimensions constructed in \cite{F4SUGRA1} and \cite{F4SUGRA2}. Although the construction given in \cite{F4SUGRA1} and \cite{F4SUGRA2} mainly focuses on compact gauge groups of the form $SU(2)\times G_c$, the formulation is valid for other types of gauge groups with appropriate modifications in some of the formulae, see also \cite{AdS6_Jan}. 
\\
\indent The matter-coupled $F(4)$ gauged supergravity consists of the supergravity and vector multiplets with the following field contents 
\begin{equation}
\left(e^{\hat{\mu}}_\mu,\psi^A_\mu, A^\alpha_\mu, B_{\mu\nu}, \chi^A,
\sigma\right)\qquad \textrm{and}\qquad (A_\mu,\lambda_A,\phi^\alpha)^I, \qquad I=1, 2, \ldots, n
\end{equation}
with $n$ being the number of vector multiplets. We use the metric signature $(-+++++)$ with space-time and flat indices denoted respectively by $\mu,\nu=0,\ldots ,5$ and $\hat{\mu},\hat{\nu}=0,\ldots, 5$. Indices $A,B,\ldots =1,2$ correspond to the fundamental representation of $SU(2)_R\sim USp(2)_R\sim SO(3)_R$ R-symmetry while indices $\alpha,\beta=(0,r)$ correspond to singlet, $0$, and adjoint indices, $r,s,\ldots =1,2,3$, of $SU(2)_R$. 
\\
\indent The bosonic fields are given by the graviton $e^{\hat{\mu}}_\mu$, a two-form field $B_{\mu\nu}$, $n+4$ vector fields $A^\Lambda=(A^\alpha_\mu,A^I_\mu)$, $\Lambda=(\alpha,I)$, and $1+4n$ scalars parametrizing the scalar manifold 
\begin{equation}
\mathbb{R}^+\times SO(4,n)/SO(4)\times SO(n)\, .
\end{equation}
The $\mathbb{R}^+$ factor corresponds to the dilaton $\sigma$ while the $4n$ scalars from the vector multiplets $\phi^{\alpha I}$ are encoded in the $SO(4,n)/SO(4)\times SO(n)$ coset representative
\begin{equation}
{L^\Lambda}_{\ul{\Sigma}}=({L^\Lambda}_\alpha,{L^\Lambda}_I).
\end{equation}
The inverse of ${L^\Lambda}_{\ul{\Sigma}}$ will be denoted by ${(L^{-1})^{\ul{\Lambda}}}_\Sigma=({(L^{-1})^{\alpha}}_\Sigma,{(L^{-1})^{I}}_\Sigma)$. Finally, the fermionic fields are given by two gravitini $\psi^A_\mu$, two spin-$\frac{1}{2}$ fields $\chi^A$, and $2n$ gaugini $\lambda^I_A$. 
\\
\indent As shown in \cite{F4SUGRA1} and \cite{F4SUGRA2}, there are two types of supersymmetric deformations namely the usual gauging of a subgroup $G_0$ of the global symmetry $\mathbb{R}^+\times SO(4,n)$ and a massive deformation leading to a mass term for the two-from field. Any consistent gauging is implemented by the gauge structure constant ${f_{\Lambda\Sigma}}^\Delta$ appearing in the $G_0$ algebra
\begin{equation}
\left[T_\Lambda,T_\Sigma\right]={f_{\Lambda\Sigma}}^\Delta T_\Delta
\end{equation}
with $T_\Lambda$ being gauge generators. In particular, the gauge covariant derivative of the coset representative is given by
\begin{equation}
\nabla_\mu {L^\Sigma}_{\ul{\Delta}}=\pd_\mu {L^\Sigma}_{\ul{\Delta}}-f^{\phantom{\Gamma}\Sigma}_{\Gamma\phantom{\Sigma}\Lambda} A^\Gamma_\mu {L^\Lambda}_{\ul{\Delta}}
\end{equation}
with $f^{\phantom{\Gamma}\Sigma}_{\Gamma\phantom{\Sigma}\Lambda}$ related to ${f_{\Gamma\Sigma}}^\Lambda$ via $f^{\phantom{\Gamma}\Sigma}_{\Gamma\phantom{\Sigma}\Lambda}={f_{\Gamma\Delta}}^\Pi\eta_{\Pi\Lambda}\eta^{\Sigma\Delta}$. $SO(4,n)$ indices $\Lambda, \Sigma,\ldots$ are raised and lowered by $SO(4,n)$ invariant tensor
\begin{equation}
\eta^{\Lambda\Sigma}=\eta_{\Lambda\Sigma}=(\delta_{\alpha\beta},-\delta_{IJ}).
\end{equation} 
We also emphasize that supersymmetry requires ${f_{\Lambda\Sigma}}^\Gamma$ to satisfy $f_{\Lambda\Gamma\Sigma}={f_{\Lambda\Gamma}}^\Delta\eta_{\Delta\Sigma}=f_{[\Lambda\Gamma\Sigma]}$.
\\
\indent In this paper, we will consider only vacua, holographic RG flows and supersymmetric Janus solutions which involve only the metric and scalars. Accordingly, we will set all the other fields to zero in order to further simplify various subsequent formulae. With only scalars and the metric non-vanishing, the bosonic Lagrangian of the matter-coupled $F(4)$ gauged supergravity can be written as 
\begin{eqnarray}
e^{-1}\mathcal{L}&=&\frac{1}{4}R-\pd_\mu \sigma\pd^\mu \sigma
-\frac{1}{4}P^{I\alpha}_\mu P^{\mu}_{I\alpha}-V\label{Lar}
\end{eqnarray}
with $e=\sqrt{-g}$. The kinetic term for scalars $\phi^{\alpha I}$ is written in terms of the vielbein $P^{I\alpha}_\mu$ on $SO(4,n)/SO(4)\times SO(n)$ coset. This can be obtained from the left-invariant 1-form
\begin{equation}
{\Omega^{\ul{\Lambda}}}_{\ul{\Sigma}}=
{(L^{-1})^{\ul{\Lambda}}}_{\Pi}\nabla {L^\Pi}_{\ul{\Sigma}}
\end{equation}
by the following identification  
\begin{equation}
P^I_{\phantom{s}\alpha}=(P^I_{\phantom{a}0},P^I_{\phantom{a}r})=(\Omega^I_{\phantom{a}0},\Omega^I_{\phantom{a}r}).
\end{equation}
The other components $(\Omega^{rs},\Omega^{r0},\Omega^{IJ})$ lead to the $SO(4)\times SO(n)$ composite connections.
\\
\indent The scalar potential is given by
\begin{eqnarray}
V&=&-e^{2\sigma}\left[\frac{1}{36}A^2+\frac{1}{4}B^iB_i+\frac{1}{4}\left(C^I_{\phantom{s}t}C_{It}+4D^I_{\phantom{s}t}D_{It}\right)\right]
+m^2e^{-6\sigma}\mc{N}_{00}\nonumber \\
& &-me^{-2\sigma}\left[\frac{2}{3}AL_{00}-2B^iL_{0i}\right]
\end{eqnarray}
with $\mc{N}_{00}$ being the $00$ component of $\mc{N}_{\Lambda\Sigma}$ defined as
\begin{eqnarray}
\mc{N}_{\Lambda\Sigma}=L_{\Lambda\alpha}{(L^{-1})^\alpha}_\Sigma-L_{\Lambda I}{(L^{-1})^I}_\Sigma=(\eta L L^T\eta)_{\Lambda\Sigma}\, .
\end{eqnarray}
It should be noted that this implies $\mc{N}_{\Lambda\Sigma}=\mc{N}_{\Sigma\Lambda}$. The fermion-shift matrices are defined as follows
\begin{eqnarray}
A&=&\epsilon^{rst}K_{rst},\qquad B^i=\epsilon^{ijk}K_{jk0},\\
C^{\phantom{ts}t}_I&=&\epsilon^{trs}K_{rIs},\qquad D_{It}=K_{0It}
\end{eqnarray}
where the ``dressed'' or ``boosted'' structure constants are given by
\begin{eqnarray}
K_{rs\alpha}&=&f^{\phantom{\Lambda}\Gamma}_{\Lambda\phantom{\Gamma}\Sigma}{L^\Lambda}_r(L^{-1})_{s\Gamma}{L^\Sigma}_\alpha,\nonumber \\
K_{\alpha It}&=&f^{\phantom{\Lambda}\Gamma}_{\Lambda\phantom{\Gamma}\Sigma}{L^\Lambda}_\alpha (L^{-1})_{I\Gamma}{L^\Sigma}_t
\end{eqnarray}
with $\alpha=(0,r)$. We also note that the gauge coupling constant is included in the structure constants $f^{\phantom{\Lambda}\Gamma}_{\Lambda\phantom{\Gamma}\Sigma}$. The constant $m$ corresponds to the aforementioned massive deformation.
\\
\indent Finally, supersymmetry transformations for fermionic fields are given by
\begin{eqnarray}
\delta\psi_{\mu
A}&=&D_\mu\epsilon_A-\frac{1}{24}\left(Ae^\sigma+6me^{-3\sigma}(L^{-1})_{00}\right)\epsilon_{AB}\gamma_\mu\epsilon^B\nonumber
\\
& &-\frac{1}{8}
\left(B_te^\sigma-2me^{-3\sigma}(L^{-1})_{t0}\right)\gamma^7\sigma^t_{AB}\gamma_\mu\epsilon^B,\label{delta_psi}\\
\delta\chi_A&=&\frac{1}{2}\gamma^\mu\pd_\mu\sigma\epsilon_{AB}\epsilon^B+\frac{1}{24}
\left[Ae^\sigma-18me^{-3\sigma}(L^{-1})_{00}\right]\epsilon_{AB}\epsilon^B\nonumber
\\
& &-\frac{1}{8}
\left[B_te^\sigma+6me^{-3\sigma}(L^{-1})_{t0}\right]\gamma^7\sigma^t_{AB}\epsilon^B,\label{delta_chi}\\
\delta
\lambda^{I}_A&=&P^I_{ri}\gamma^\mu\pd_\mu\phi^i\sigma^{r}_{\phantom{s}AB}\epsilon^B+P^I_{0i}
\gamma^7\gamma^\mu\pd_\mu\phi^i\epsilon_{AB}\epsilon^B-\left(2i\gamma^7D^I_{\phantom{s}t}+C^I_{\phantom{s}t}\right)
e^\sigma\sigma^t_{AB}\epsilon^B \nonumber
\\
& &+2me^{-3\sigma}(L^{-1})^I_{\phantom{ss}0}
\gamma^7\epsilon_{AB}\epsilon^B\label{delta_lambda}
\end{eqnarray}
with $\sigma^{rA}_{\phantom{sd}B}$ being Pauli matrices and $\epsilon_{AB}=-\epsilon_{BA}$. The $SU(2)$ fundamental indices $A,B,\ldots$ are raised and lowered by $\epsilon^{AB}$ and $\epsilon_{AB}$ with the convention $T^A=\epsilon^{AB}T_B$ and $T_A=T^B\epsilon_{BA}$. 
\\
\indent The covariant derivative of $\epsilon_A$ is defined by
\begin{equation}
D_\mu \epsilon_A=\pd_\mu
\epsilon_A+\frac{1}{4}\omega_\mu^{ab}\gamma_{ab}\epsilon_A+\frac{i}{2}\sigma^r_{AB}
\left[\frac{1}{2}\epsilon^{rst}\Omega_{\mu st}-i\gamma_7
\Omega_{\mu r0}\right]\epsilon^B\, .
\end{equation}
Gamma matrices $\gamma^a$ satisfy the Clifford algebra
\begin{equation}
\{\gamma^a,\gamma^b\}=2\eta^{ab},\qquad
\eta^{ab}=\textrm{diag}(-1,1,1,1,1,1).
\end{equation}
The chirality matrix is defined by $\gamma_7=i\gamma^0\gamma^1\gamma^2\gamma^3\gamma^4\gamma^5$ with
$\gamma_7^2=-\mathbf{1}$. 
\section{$ISO(3)\times U(1)$ gauge group and vacua}\label{Vacua}
We now consider $F(4)$ gauged supergravity coupled to $n=4$ vector multiplets with $ISO(3)\times U(1)$ gauge group. This gauged supergravity has been shown to arise from a consistent truncation of type IIB theory on $S^2\times \Sigma$ in \cite{Henning_Malek_AdS7_6}. In this case, the global symmetry is given by $\mathbb{R}^+\times SO(4,4)$. As previously mentioned, the $ISO(3)\sim SO(3)\ltimes \mathbf{R}^3$ factor is embedded in $SO(4,3)\subset SO(4,4)$ while the $U(1)$ factor is the abelian gauge symmetry associated with the fourth vector multiplet. We also point out that the other fields are not charged under this $U(1)$ factor as in the ungauged supergravity.
\\
\indent The embedding of $ISO(3)$ is given by the following structure constants
\begin{eqnarray}
f_{rst}=g\epsilon_{rst},\qquad f_{r\bar{s}\bar{t}}=-g\epsilon_{r\bar{s}\bar{t}},\qquad f_{\bar{r}\bar{s}\bar{t}}=-2g\epsilon_{\bar{r}\bar{s}\bar{t}}
\end{eqnarray}
in which we have split the indices $I,J=1,2,3,4$ as $I=(\bar{r},4)$ with $\bar{r},\bar{s}=1,2,3$. The gauge generators are defined by $X_\Lambda=f_{\Lambda\Sigma\Gamma}T^{\Sigma\Gamma}$ with $T^{\Sigma\Gamma}=-T^{\Gamma\Sigma}$ being $SO(4,4)$ generators given by $(T^{\Lambda\Sigma})_\Gamma^{\phantom{\Gamma}\Delta}=2\delta^{[\Lambda}_\Gamma \eta^{\Sigma]\Delta}$. From the structure constants, we can straightforwardly verify that $X_0=0$ and $X_r$ generate the compact $SO(3)$ subgroup while $X_r-X_{\bar{r}}$ form a set of three commuting generators in $\mathbf{R}^3$ and transform as a vector or adjoint representation of $SO(3)$. It should also be noted that in general, the gauge coupling constant $g$ and the mass parameter $m$ of the massive deformation are two independent parameters. However, the consistent truncations found in \cite{Henning_Malek_AdS7_6} require the relation $g=3m$. This is very similar to the embedding of pure $F(4)$ gauged supergravity \cite{F4_Romans} in massive type IIA theory using the truncation given in \cite{Massive_IIA_onS4}. In that case, only the gauged supergravity with $g=3m$ (called $N=4^+$ in \cite{F4_Romans}) can be uplifted to ten dimensions. 
\\
\indent We are now in a position to compute the scalar potential and look for possible vacua. We follow the usual approach to simplify the computation by considering a truncation invariant under a certain subgroup of the gauge group. In particular, we will consider $SO(3)\subset ISO(3)$ invariant scalars. Including the $U(1)$ factor, these scalars are invariant under $SO(3)\times U(1)\subset ISO(3)\times U(1)$ symmetry. To give an explicit parametrization of $SO(4,4)/SO(4)\times SO(4)$ coset, we introduce the following $GL(8,\mathbb{R})$ matrices
\begin{equation}
(e^{\Lambda \Sigma})_{\Gamma \Pi}=\delta^\Lambda_{
\Gamma}\delta^\Sigma_{\Pi},\qquad \Lambda, \Sigma,\Gamma,
\Pi=0,\ldots ,7\, .
\end{equation}
The non-compact generators can be identified with
\begin{eqnarray}
Y_{\alpha I}=e^{\alpha,I+3}+e^{I+3,\alpha}\, .
\end{eqnarray}
\indent We first recall that the $16$ scalars transform under the compact $SO(4)\times SO(4)\subset SO(4,4)$ subgroup as $(\mathbf{4},\mathbf{4})$. As previously mentioned, the $SO(3)$ residual symmetry is generated by $X_r$. This $SO(3)$ corresponds to a diagonal subgroup of $SO(3)\times SO(3)\subset SO(4)\times SO(4)$ in which each $SO(3)$ factor is a subgroup of each $SO(4)$ with the embedding described by the following decomposition $\mathbf{4}\rightarrow \mathbf{1}+\mathbf{3}$. This implies the branching
\begin{equation}
(\mathbf{4},\mathbf{4})\rightarrow (\mathbf{1},\mathbf{1})+(\mathbf{1},\mathbf{3})+(\mathbf{3},\mathbf{1})+(\mathbf{3},\mathbf{3})
\end{equation} 
under $SO(4)\times SO(4)\rightarrow SO(3)\times SO(3)$. By further decomposing under $SO(3)\sim [SO(3)\times SO(3)]_{\textrm{diag}}$, we end up with
\begin{equation}
(\mathbf{4},\mathbf{4})\rightarrow \mathbf{1}'+\mathbf{3}_{\textrm{v}}+\mathbf{3}'_{\mathbf{v}}+\mathbf{1}+\mathbf{3}_{\textrm{Adj}}+\mathbf{5}\, .\label{SO3_decom}
\end{equation}
For convenience, we have distinguished the two singlets from $(\mathbf{1},\mathbf{1})$ and from $(\mathbf{3},\mathbf{3})$ as $\mathbf{1}'$ and $\mathbf{1}$. Similarly, $\mathbf{3}_{\textrm{v}}$ and $\mathbf{3}'_{\textrm{v}}$ denote the $SO(3)$ vector representations arising from $(\mathbf{1},\mathbf{3})$ and $(\mathbf{3},\mathbf{1})$, respectively.   
\\
\indent In terms of the above $SO(4,4)$ generators, the two singlet scalars $\mathbf{1}$ and $\mathbf{1}'$ correspond respectively to the following non-compact generators
\begin{equation}
\hat{Y}_1=Y_{11}+Y_{22}+Y_{33}\qquad \textrm{and}\qquad \hat{Y}_2=Y_{04}\, .
\end{equation}
The coset representative can then be written as
\begin{equation}
L=e^{\phi \hat{Y}_1}e^{\varphi \hat{Y}_2}\, .\label{L_SO3}
\end{equation}
Together with the dilaton $\sigma$, there are three singlet scalars under $SO(3)$. The scalar potential restricted to these singlets is given by
\begin{eqnarray}
V&=&\frac{1}{2}\left[g^2e^{2\sigma+2\phi}(e^{4\phi}-3)+m^2e^{-6\sigma-2\varphi}(1+e^{4\varphi})\right]\nonumber \\
& &+gme^{-2\sigma+\phi-\varphi}(1+e^{2\varphi})(e^{2\phi}-3).\label{V_SO3}
\end{eqnarray}
We find three critical points from this potential given as follows:
\begin{itemize}
\item I.
\begin{equation}
\phi=\varphi=0,\qquad \sigma=\frac{1}{4}\ln\left[\frac{3m}{g}\right],\qquad V_0=-\frac{20g^2}{3}\sqrt{\frac{m}{3g}}\, .
\end{equation}
\item II.
\begin{equation}
\phi=\varphi=0\qquad \sigma=\frac{1}{4}\ln\left[\frac{m}{g}\right],\qquad V_0=-4g^2\sqrt{\frac{m}{g}}\, .
\end{equation}
\item III.
\begin{eqnarray}
\varphi&=&0,\qquad \phi=\frac{1}{2}\ln\left[\frac{7+4\sqrt{7}}{3}\right],\nonumber \\
\sigma&=&-\frac{1}{4}\ln\left[-\frac{g(5+2\sqrt{7})}{3m}\sqrt{\frac{7+4\sqrt{7}}{3}}\right],\nonumber \\
V_0&=&-\frac{4gm(7+4\sqrt{7})^{\frac{7}{4}}}{3^{\frac{9}{4}}}\sqrt{-\frac{g}{m(5+2\sqrt{7})}}\, .
\end{eqnarray}
\end{itemize}
$V_0$ denotes the cosmological constant given by the value of the scalar potential at the critical point.
\\
\indent Critical point I is a supersymmetric $AdS_6$ vacuum as will be seen in the next section. We can also choose $g=3m$ or equivalently shift the value of the dilaton at this vacuum to zero. This is precisely the relation arising from the truncations of type IIB theory as previously mentioned. All scalar masses at this critical point is given in table \ref{table1}. The $AdS_6$ radius $L$ is given by $L^2=-\frac{5}{V_0}=\frac{1}{4m^2}$ for $g=3m$. The scalars are denoted by their $SO(3)$ representations. To make the identification clearer, we have distinguished the same representations as follows: $\mathbf{1}$, $\mathbf{1}'$ and $\mathbf{1}_\sigma$ correspond to $SO(3)$ singlet scalars $\phi$, $\varphi$ and $\sigma$, respectively. $\mathbf{3}_\textrm{v}$ and $\mathbf{3}'_\textrm{v}$ refer respectively to scalars corresponding to non-compact generators $Y_{0\bar{r}}$ and $Y_{r4}$. The three massless scalars $\mathbf{3}_\textrm{Adj}$ in the adjoint representation of $SO(3)$ are Goldstone bosons of the symmetry breaking $ISO(3)\rightarrow SO(3)$ at the vacuum.  
\begin{table}[!htbp]
\renewcommand{\arraystretch}{1.5}
\centering
\begin{tabular}{|c|c|c|}
\hline
Scalars & $m^2L^2$ & $\Delta $\\
\hline
$\mathbf{1}_\sigma$ & $-6$ & $3$\\ \hline
$\mathbf{1}$ & $24$ & $8$ \\ \hline
$\mathbf{3_\textrm{Adj}}$ & $0$  & $5$\\ \hline
$\mathbf{5}$ & $6$ & $6$\\ \hline
$\mathbf{3}_\textrm{v}$ & $14$ &  $7$\\ \hline
$\mathbf{1}'$ & $-4$  & $4$\\ \hline
$\mathbf{3}'_\textrm{v}$ & $-6$  &  $3$\\ \hline
\end{tabular}
\caption{Scalar masses at the supersymmetric $AdS_6$ vacuum, critical point I, and the corresponding dimensions of the dual operators.}\label{table1}
\end{table}
This $AdS_6$ vacuum should be dual to an $N=2$ SCFT in five dimensions. In the table, we have also included the dimensions of the operators dual to all of the scalars via the relation
\begin{equation}
m^2L^2=\Delta(\Delta-5).
\end{equation}
\indent Critical point II is also an $AdS_6$ vacuum but breaks all supersymmetry. However, this critical point is unstable as can be seen from the scalar masses given in table \ref{table2}. It is useful to note the BF bound in six-dimensional gauged supergravity $m^2L^2=-\frac{25}{4}$.
\begin{table}[!htbp]
\renewcommand{\arraystretch}{1.5}
\centering
\begin{tabular}{|c|c|c|}
\hline
Scalars & $m^2L^2$ \\
\hline
$\mathbf{1}_\sigma$ & $10$ \\ \hline
$\mathbf{1}$ & $20$ \\ \hline
$\mathbf{3_\textrm{Adj}}$ & $0$  \\ \hline
$\mathbf{5}$ & $-10$ \\ \hline
$\mathbf{3}_\textrm{v}$ & $10$ \\ \hline
$\mathbf{1}'$ & $0$  \\ \hline
$\mathbf{3}'_\textrm{v}$ & $-10$  \\ \hline
\end{tabular}
\caption{Scalar masses at the non-supersymmetric $AdS_6$ vacuum, critical point II.}\label{table2}
\end{table}
\\
\indent Finally, critical point III is a $dS_6$ vacuum. The reality of $\sigma$ at this critical point requires $gm<0$ which in turn implies that $V_0>0$. It should be noted that this $dS_6$ vacuum cannot be embedded in type IIB theory since the consistent truncations are only possible for $g=3m$. This is also in agreement with the no-go theorem found in \cite{maldacena_nogo}. The $dS_6$ vacuum is however unstable with the scalar masses given in table {\ref{table3}}. 
\begin{table}[!htbp]
\renewcommand{\arraystretch}{1.5}
\centering
\begin{tabular}{|c|c|c|}
\hline
Scalars & $m^2L^2$ \\
\hline
$\mathbf{1}_\sigma$, $\mathbf{1}$ & $-\frac{5}{7}(77+36\sqrt{7}\pm \sqrt{22057+7896\sqrt{7}})$ \\ \hline
$\mathbf{3_\textrm{Adj}}$ & $0$  \\ \hline
$\mathbf{5}$ & $-10-\frac{80}{\sqrt{7}}$ \\ \hline
$\mathbf{3}_\textrm{v}$ & $-50-\frac{120}{\sqrt{7}}$ \\ \hline
$\mathbf{1}'$ & $-20-\frac{40}{\sqrt{7}}$  \\ \hline
$\mathbf{3}'_\textrm{v}$ & $10-\frac{40}{\sqrt{7}}$  \\ \hline
\end{tabular}
\caption{Scalar masses at the $dS_6$ vacuum, critical point III. The first two masses correspond to two different linear combinations of $\mathbf{1}_\sigma$ and $\mathbf{1}$.}\label{table3}
\end{table}
\\
\indent We also note that there are no other supersymmetric $AdS_6$ with smaller residual symmetry such as $SO(2)\subset SO(3)$ since the existence of these vacua requires at least $SO(3)$ unbroken symmetry, see more detail in \cite{AdS6_Jan}. 
\section{Holographic RG flows}\label{RG_flows}
In this section, we consider supersymmetric solutions describing holographic RG flows in the $N=2$ five-dimensional SCFT  dual to the superstmmetric $AdS_6$ vacuum identified in the previous section. The metric ansatz is given by the standard domain wall
\begin{equation}
ds^2=e^{2A(r)}dx^2_{1,4}+dr^2
\end{equation}
with $dx^2_{1,4}$ being the metric on five-dimensional Minkowski space. The metric warp factor $A(r)$ and all non-vanishing scalar fields are only functions of the radial coordinate $r$. 
\\
\indent To find supersymmetric solutions, we solve the corresponding BPS equations arising from the supersymmetry transformations of fermionic fields. For non-constant  scalars, we will also need to impose a projection
\begin{equation}
\gamma_r\epsilon_A=-\epsilon_A\, .\label{RG_flow_proj}
\end{equation}
The sign choice is chosen to identify the supersymmetric $AdS_6$ vacuum with the limit $r\rightarrow \infty$ at which $A\sim \frac{r}{L}$ for the $AdS_6$ radius $L$. Since there is only one supersymmetric $AdS_6$ vacuum, all the supersymmetric holographic RG flows are essentially RG flows to non-conformal phases.  

\subsection{RG flows from $SO(3)$ invariant sector}
We first consider RG flow solutions with $SO(3)$ singlet scalars given by the coset representative \eqref{L_SO3}. After imposing the projection condition \eqref{RG_flow_proj}, we find the following BPS equations 
\begin{eqnarray}
& &A'=W=\frac{1}{4}\left[ge^\sigma(3e^{\phi}-e^{3\phi})+me^{-3\sigma}(e^{-\varphi}+e^\varphi)\right],\\
& &\phi'=-\frac{4}{3}\frac{\pd W}{\pd\phi}=2ge^{\sigma+2\phi}\sinh\phi,\\
& &\varphi'=-4\frac{\pd W}{\pd \varphi}=-2me^{-3\sigma}\sinh\varphi,\\
& &\sigma'=-\frac{\pd W}{\pd \sigma}=-\frac{1}{4}\left[ge^\sigma(3e^\phi-e^{3\phi})-3me^{-3\sigma}(e^\varphi+e^{-\varphi})\right].
\end{eqnarray}
Throughout the paper, we use $'$ to denote $r$-derivatives. In these equations, we have also introduced the superpotential $W$ in terms of which the scalar potential \eqref{V_SO3} can be written as
\begin{equation}
V=\left(\frac{\pd W}{\pd \sigma}\right)^2+\frac{4}{3}\left(\frac{\pd W}{\pd \phi}\right)^2+4\left(\frac{\pd W}{\pd \varphi}\right)^2-5W^2\, .
\end{equation}
It can be readily verified that the BPS equations are compatible with the corresponding field equations. We also note that the BPS equations admit a fixed point solution at which all scalars are constant, $\phi'=\varphi'=\sigma'=0$, and $A=\frac{r}{L}$ with $L=\frac{1}{2m}$ for $g=3m$. This is the supersymmetric $AdS_6$ vacuum denoted by critical point I in the previous section.

\subsubsection{An RG flow driven by relevant operators} 
We first consider a simpler solution with $\phi=0$ which is a consistent truncation of the above BPS equations. According to the scalar masses given in table \ref{table1}, $\phi$ is dual to an irrelevant operator of dimension $\Delta=8$. Therefore, solutions with $\phi=0$ holographically describe RG flows driven solely by relevant operators dual to $\varphi$ and $\sigma$. 
\\
\indent By considering $A$ and $\sigma$ as functions of $\varphi$, we can solve for $A(\varphi)$ and $\sigma(\varphi)$ as follows
\begin{eqnarray}
& &\sigma=-\frac{1}{4}\ln\left[\frac{g}{6m}(3\cosh\varphi-\cosh3\varphi+2\sigma_0\sinh^3\varphi)\right],\nonumber \\
& &A=\frac{1}{4}\ln\left[6\cosh\varphi-2\cosh3\varphi+4\sigma_0\sinh^3\varphi\right]-\ln\sinh\varphi
\end{eqnarray} 
with $\sigma_0$ being an integration constant. An additive integration constant for $A$ is removed by rescaling the coordinates on $dx^2_{1,4}$. Finally, by changing to a new radial coordinate $\rho$ via $\frac{d\rho}{dr}=e^{-3\sigma}$, we find the solution for $\varphi$ given by
\begin{equation}
\varphi=\ln\left[\frac{1+e^{-2m(\rho-\rho_0)}}{1-e^{-2m(\rho-\rho_0)}}\right]
\end{equation}
with another integration constant $\rho_0$. This constant however can be removed by shifting the coordinate $\rho$. We also note that the solution approaches the supersymmetric $AdS_6$ vacuum with $\sigma\sim \frac{1}{4}\ln\left[\frac{3m}{g}\right]$ and $\varphi\sim 0$ for any value of $\sigma_0$.
\\
\indent We now consider asymptotic behaviors of the solution. For simplicity, we will also set $g=3m$. As $\rho\rightarrow \infty$, we find $\varphi\sim\sigma\sim 0$, $\rho\sim r$ and $A\sim \frac{r}{L}$ with $L=\frac{1}{2m}$. The behavior of scalars near the $AdS_6$ vacuum is given by 
\begin{eqnarray}
& &\varphi\sim e^{2mr}\sim e^{-\frac{r}{L}},\nonumber \\ 
& &\sigma\sim \frac{3}{8}e^{-4mr}-\frac{\sigma_0}{4}e^{-6mr}\sim \frac{3}{8}e^{-\frac{2r}{L}}-\frac{\sigma_0}{4}e^{-\frac{3r}{L}}
\end{eqnarray}
in accordance with the fact that $\varphi$ and $\sigma$ are dual to operators of dimensions $\Delta=4$ and $\Delta=3$, respectively, see table \ref{table1}. We also note that the source terms for both operators are turned on. The asymptotic behavior also indicates the presence of a vacuum expectation value for the operator dual to $\sigma$ governed by the constant $\sigma_0$.  
\\
\indent For $\rho\rightarrow \rho_0$, the solution becomes singular with 
\begin{eqnarray}
\varphi&\sim& -\ln\left[2m(\rho-\rho_0)\right],\nonumber \\
\sigma&\sim& -\frac{1}{4}\ln\left[\frac{(\sigma_0-2)}{32m^3}(\rho-\rho_0)^{-3}+\frac{3(2-\sigma_0)}{8m}(\rho-\rho_0)^{-1}+\frac{3}{2}m(2+\sigma_0)(\rho-\rho_0)\right],\quad\nonumber \\
A&\sim& \frac{1}{4}\ln\left[\frac{\sigma_0-2}{16m^3}(\rho-\rho_0)^{-3}-\frac{3(2-\sigma_0)(\rho-\rho_0)^{-1}}{4m}+\frac{3}{2}m(2+\sigma_0)(\rho-\rho_0)\right]\nonumber \\
& &+\ln \left[2m(\rho-\rho_0)\right].
\end{eqnarray}
There are two possibilities depending on whether $\sigma_0=2$ or $\sigma_0\neq 2$. For $\sigma_0=2$, we find 
\begin{equation}
\sigma\sim-\frac{1}{4}\ln(\rho-\rho_0)\qquad \textrm{and}\qquad A\sim \frac{5}{4}\ln(\rho-\rho_0).
\end{equation}
The scalar potential near this singularity is bounded from above with $V\rightarrow -\infty$. Therefore, according to the criterion given in \cite{Gubser_singularity}, this singularity is acceptable and should lead to a gravity dual of a non-conformal phase of the $N=2$ SCFT in five dimensions dual to the supersymmetric $AdS_6$ vacuum. 
\\
\indent On the other hand, for $\sigma_0\neq 2$, we find
\begin{equation}
\sigma\sim \frac{3}{4}\ln(\rho-\rho_0),\qquad A\sim \frac{1}{4}\ln(\rho-\rho_0),\qquad V\rightarrow +\infty\, .
\end{equation}
In this case, the singularity is unphysical. It would be useful to have a complete truncation ansatz to uplift the solution to type IIB theory and verify whether the singularity is resolved in ten dimensions.

\subsubsection{An RG flow driven by relevant and irrelevant operators}  
We now consider solutions with all $SO(3)$ singlet scalars non-vanishing. To solve the BPS equations, we first take the following combinations
\begin{eqnarray}
& &\frac{dA}{d\varphi}+\frac{d\sigma}{d\varphi}=\frac{1+e^{2\varphi}}{1-e^{2\varphi}},\\
& &\frac{d\sigma}{d\phi}-3\frac{dA}{d\phi}=\frac{e^{2\phi}-3}{e^{2\phi}-1}\, .
\end{eqnarray}
We then find the solutions for $A$ and $\sigma$ of the form
\begin{eqnarray}
& &A=\frac{1}{4}(\varphi-3\phi)-\frac{1}{4}\ln(1-e^{2\varphi})+\frac{1}{4}\ln(1-e^{2\phi}),\\
& &\sigma=\frac{3}{4}(\phi+\varphi)-\frac{3}{4}\ln(1-e^{2\varphi})-\frac{1}{4}\ln\left[\frac{1-e^{2\phi}}{16}\right]
\end{eqnarray}
in which we have removed an additive integration constant for $A$ and choosing appropriate value of the integration constant for $\sigma$ such that the solution approach the $AdS_6$ vacuum as $\phi\rightarrow 0$ and $\varphi\rightarrow 0$.
\\
\indent Using the solution for $\sigma$ in the remaining two equations, we finally obtain the solutions for $\phi$ and $\varphi$ given by
\begin{eqnarray}
& &\phi=-\frac{1}{2}\ln\left[C_0-\frac{16(3e^{2\varphi}-1)}{(e^{2\varphi}-1)^3}\right],\\
& &\varphi=\ln\left[\frac{1+e^{-2m(\rho-\rho_0)}}{1-e^{-2m(\rho-\rho_0)}}\right]
\end{eqnarray}
with $\rho$ defined by $\frac{d\rho}{dr}=e^{-3\sigma}$ and $C_0$ and $\rho_0$ being integration constants. For convenience, we have also set $g=3m$ in these solutions. As in the previous case, the solution is singular for $\rho\rightarrow \rho_0$. Near the singularity, the solution becomes
\begin{equation}
\varphi\sim -\ln(\rho-\rho_0)
\end{equation}
which leads to
\begin{eqnarray}
\phi\sim -\ln(\rho-\rho_0),\qquad \sigma\sim \frac{1}{2}\ln(\rho-\rho_0),\qquad A\sim \frac{1}{2}\ln(\rho-\rho_0),
\end{eqnarray}
for $C_0=0$, and
\begin{eqnarray}
\phi\sim \textrm{constant},\qquad \sigma\sim \frac{3}{4}\ln(\rho-\rho_0),\qquad A\sim \frac{1}{4}\ln(\rho-\rho_0),
\end{eqnarray}
for $C_0\neq 0$. Both of these give $V\rightarrow \infty$, so the singularity is unacceptable by the criterion of \cite{Gubser_singularity}.

\subsubsection{A truncation to three vector multiplets}
We finally consider another subtruncation with $\varphi=0$. In this case, the solutions will be the same as those of $F(4)$ gauged supergravity coupled to three vector multiplets with $ISO(3)$ gauge group. These solutions are interesting in the sense that the truncation ansatz for the ten-dimensional metric is given in \cite{Henning_Malek_AdS7_6}. Accordingly, we can use the criterion given in \cite{maldacena_nogo} to verify whether the singularities appearing in the six-dimensional solutions are physically acceptable in type IIB theory.    
\\
\indent Setting $\varphi=0$, we are left with the following set of BPS equations
\begin{eqnarray}
& & \phi'=ge^{\sigma+\phi}(e^{2\phi}-1),\nonumber \\
& & \sigma'=\frac{1}{4}\left[ge^{\sigma+\phi}(e^{2\phi}-3)+6me^{-3\sigma}\right],\nonumber \\
& & A'=\frac{1}{4}\left[ge^{\sigma+\phi}(3-e^{2\phi})+2me^{-3\sigma}\right].
\end{eqnarray}
By the same procedure, we find the following solution
\begin{eqnarray}
& &g(\rho-\rho_0)=e^{-\phi}+\frac{1}{2}\ln\left[\frac{1-e^{-\phi}}{1+e^{-\phi}}\right],\nonumber \\
& &\sigma=\frac{1}{4}\ln \left[\frac{e^{-\phi}(2\sigma_0e^{4\phi}-3m)}{2g(e^{2\phi}-1)}\right],\nonumber \\
& &A=\frac{1}{4}\ln(1-e^{2\phi})+\frac{1}{12}\ln (3m-2\sigma_0e^{4\phi})-\frac{13}{12}\phi
\end{eqnarray}
with $\rho$ defined by $\frac{d\rho}{dr}=e^\sigma$. We will choose the integration constant $\sigma_0=\frac{3}{2}m$ in order to make the solution approach the supersymmetric $AdS_6$ vacuum. As $\rho\rightarrow \infty$, we find, for $g=3m$,
\begin{equation}
\phi\sim e^{3mr}\sim e^{\frac{3r}{L}}\qquad \textrm{and}\qquad \sigma \sim e^{-6mr}\sim e^{-\frac{3r}{L}}
\end{equation} 
as expected. At the singularity as $\rho\rightarrow \rho_0$, we find 
\begin{eqnarray}
& &\phi\sim -\frac{1}{3}\ln [3m(\rho-\rho_0)],\qquad \sigma\sim -\frac{1}{12}\ln [3m(\rho-\rho_0)],\nonumber \\
& & A\sim \frac{1}{12}\ln [3m(\rho-\rho_0)].
\end{eqnarray}
In this case, the scalar potential becomes unbounded from above $V\rightarrow \infty$.
\\
\indent As previously mentioned, the complete truncation ansatz for the ten-dimensional metric has been given in \cite{Henning_Malek_AdS7_6}. Although this truncation ansatz is only applicable for constant scalars, we can use this result to find the $00$-component of the ten-dimensional metric to check whether the uplifted singularity is acceptable by the criterion of \cite{maldacena_nogo}. We will only present relevant ingredients for obtaining the uplifted $00$-component of the metric $\hat{g}_{00}$ and refer to \cite{Henning_Malek_AdS7_6} for more detail of the truncation. The truncation of type IIB theory on $S^2\times \Sigma$ is characterized by two holomorphic functions
\begin{equation}
f^A=-p^A+ik^A,\qquad A=1,2\, .
\end{equation}
Using the formulae in \cite{Henning_Malek_AdS7_6}, we find the $00$-component of the metric given by
\begin{equation}
\hat{g}_{00}=-\frac{12c_6R^{\frac{1}{4}}}{3^{\frac{3}{4}}|dk|^{\frac{1}{2}}}\bar{\Delta}^{\frac{1}{4}}e^{2A}
\end{equation}
with $c_6$ being a constant. The warp factor $\bar{\Delta}$ is defined by
\begin{eqnarray}
\bar{\Delta}&=&e^{4\sigma}\Xi |{m_{\bar{r}}}^r||dk|-3R|dk|^2\left\{|{m_{\bar{r}}}^r||{m_{\bar{r}}}^r-2(\lambda_{\bar{r}}.y)y^r| \phantom{\frac{1}{2}}\right.\nonumber \\
& &\left. +\frac{1}{4}\left(\epsilon_{rst}\epsilon^{\bar{r}\bar{s}\bar{t}}m_{\bar{r}}y^r{m_{\bar{s}}}^s{m_{\bar{t}}}^t\right)^2\right\}
\end{eqnarray}
with 
\begin{eqnarray}
dR&=&-p_Adk^A,\qquad |dk|=\frac{1}{2}\pd_A k_B\pd^Ak^B,\nonumber \\
\Xi&=&p_Ap_B\pd_Ck^A\pd^Cp^B,\qquad \lambda_{\bar{r}}.y={\lambda_{\bar{r}}}^ry_r
\end{eqnarray}
and $y^r$, $r=1,2,3$, being coordinates on $S^2$ satisfying $y^ry^s\delta_{rs}=1$. The notation $|{m_{\bar{r}}}^r|$ denotes the determinant of the matrix ${m_{\bar{r}}}^r$. The matrix ${m_{\bar{r}}}^\Lambda=(m_{\bar{r}},{m_{\bar{r}}}^r-{\lambda_{\bar{r}}}^r,{\lambda_{\bar{r}}}^r)$, $\Lambda=(0,r,\bar{r})$, in turn describes the $SO(4,3)/SO(4)\times SO(3)$ coset and is related to the coset representative ${L^{\Lambda}}_{\ul{\Sigma}}$ via
\begin{equation}
{m_{\bar{r}}}^\Lambda{m_{\bar{s}}}^\Sigma\eta_{\Lambda\Sigma}=-\delta_{{\bar{r}}{\bar{s}}}\qquad \textrm{and}\qquad {m_{\bar{r}}}^\Lambda{m_{\bar{s}}}^{\Sigma}\delta^{{\bar{r}}{\bar{s}}}={L^\Lambda}_\alpha{L^\Sigma}_\beta\delta^{\alpha\beta}-\eta^{\Lambda\Sigma}\, .
\end{equation}
In the present case, the coset representative is given by \eqref{L_SO3} with $\varphi=0$, we have  
\begin{equation}
m_{\bar{r}}=0,\qquad {m_{\bar{r}}}^r=(\cosh\phi+\sinh\phi)\delta^r_{\bar{r}},\qquad {\lambda_{\bar{r}}}^r=\cosh\phi \delta^r_{\bar{r}}\, .
\end{equation}
With all these, we find that near the singularity
\begin{equation}
\hat{g}_{00}\sim e^{2A+\phi}\sim e^{\frac{\phi}{2}}\rightarrow \infty\, .
\end{equation}
Therefore, the sigularity is physically unacceptable by the criterion of \cite{maldacena_nogo}. It could be interesting to see whether this singularity can be interpreted as some brane configuration in type IIB theory. However, we need a complete truncation ansatz at least with non-constant scalars in order to find the corresponding ten-dimensional solution.  

\subsection{RG flows from $SO(2)$ invariant sector}
We now consider holographic RG flow solutions with a smaller $SO(2)\subset SO(3)$ residual symmetry. By further decomposing the $SO(3)$ representations given in \eqref{SO3_decom} under the $SO(2)$ subgroup, we find six singlets with the coset representative 
\begin{equation}
L=e^{\phi_0Y_{03}}e^{\phi_1(Y_{11}+Y_{22})}e^{\phi_2Y_{33}}e^{\phi_3(Y_{12}-Y_{21})}e^{\phi_4Y_{04}}e^{\phi_5Y_{34}}\, .\label{L_SO2}
\end{equation}
However, in order to consistently truncate out all the vector fields, we need to set $\phi_3=0$. Furthermore, consistency of $\delta\lambda^I_A=0$ equations requires $\phi_0=0$. Finally, consistency of $\delta\chi_A=0$ and $\delta\psi_{A\mu}=0$ equations impose another condition of the form
\begin{equation}
\sinh\phi_4\sinh\phi_5=0\, .
\end{equation}
Accordingly, we can have solutions with either $\phi_4=0$ or $\phi_5=0$. We will see in the next section that all the three scalars $\phi_0$, $\phi_4$ and $\phi_5$ can be non-vanishing in Janus solutions described by curved domain walls with $AdS_5$ slices. In a sense, these scalars are necessary to support the curvature of the domain walls. 

\subsubsection{Solutions with $\phi_5=0$}
\begin{figure}
         \centering
               \begin{subfigure}[b]{0.45\textwidth}
                 \includegraphics[width=\textwidth]{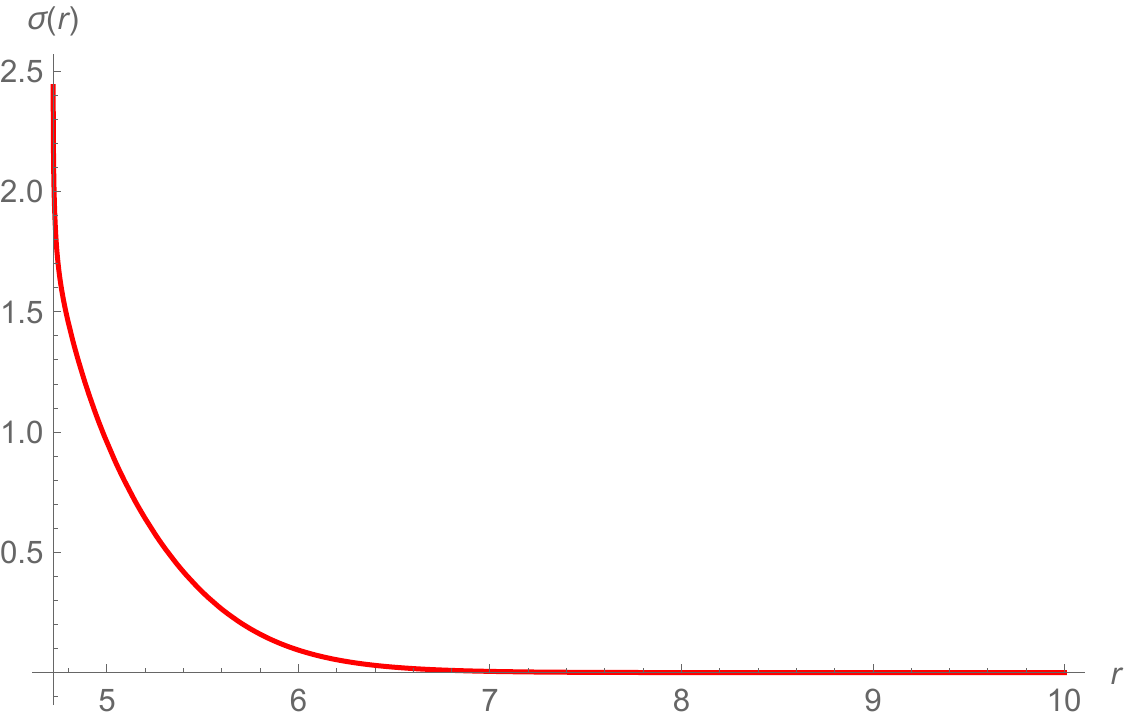}
                 \caption{Solution for $\sigma(r)$}
         \end{subfigure}
         \begin{subfigure}[b]{0.45\textwidth}
                 \includegraphics[width=\textwidth]{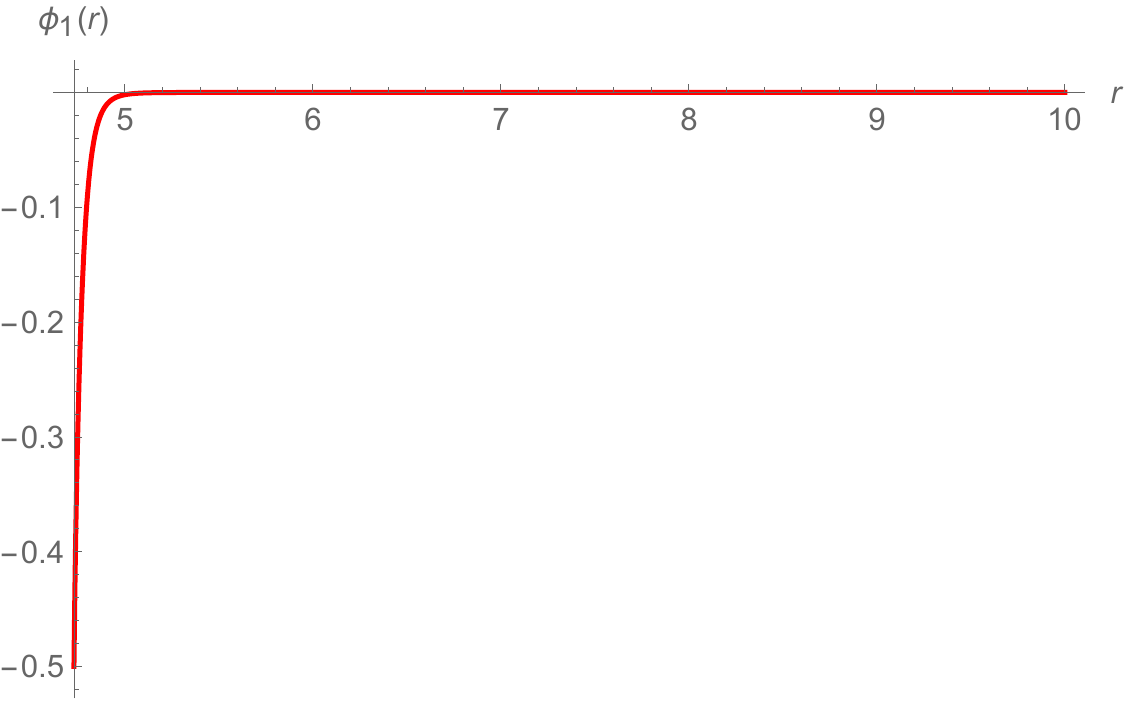}
                 \caption{Solution for $\phi_1(r)$}
         \end{subfigure}\\
          \begin{subfigure}[b]{0.45\textwidth}
                 \includegraphics[width=\textwidth]{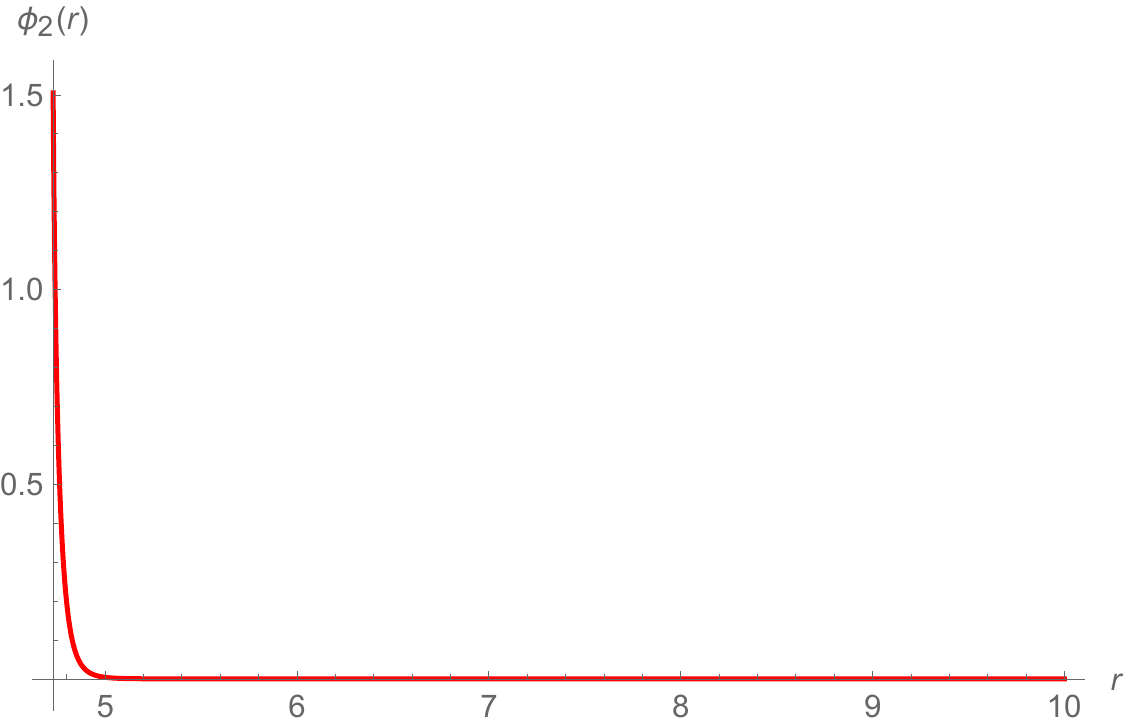}
                 \caption{Solution for $\phi_2(r)$}
         \end{subfigure}
          \begin{subfigure}[b]{0.45\textwidth}
                 \includegraphics[width=\textwidth]{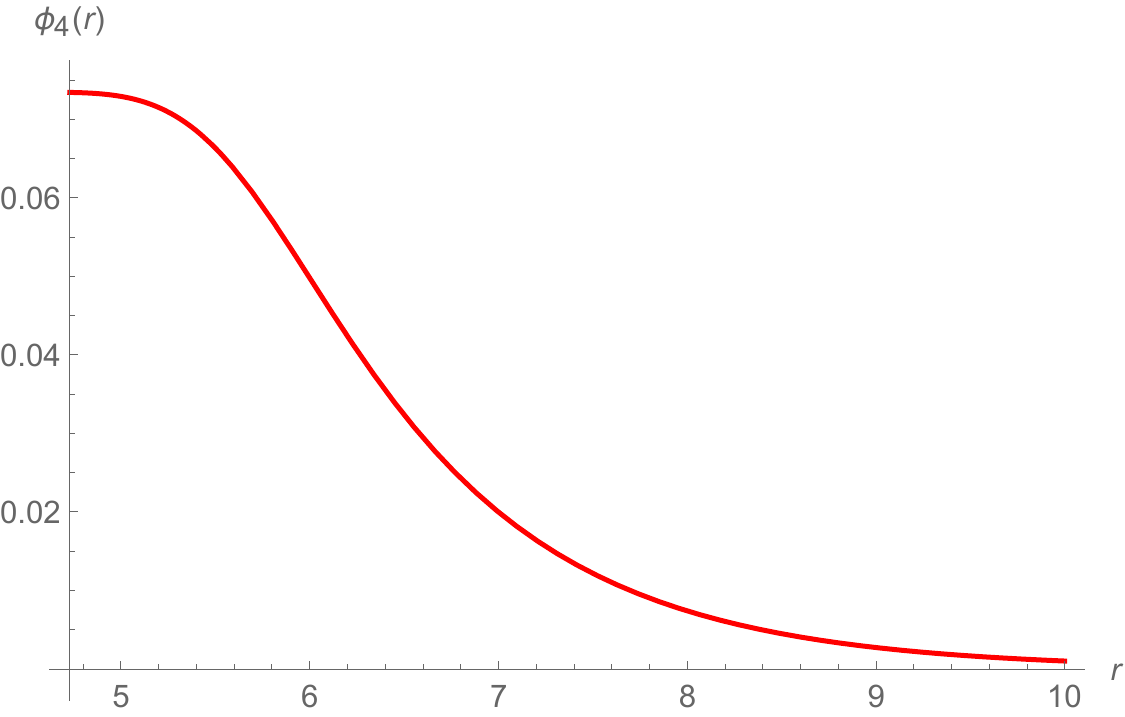}
                 \caption{Solution for $\phi_4(r)$}
         \end{subfigure}\\
         \begin{subfigure}[b]{0.45\textwidth}
                 \includegraphics[width=\textwidth]{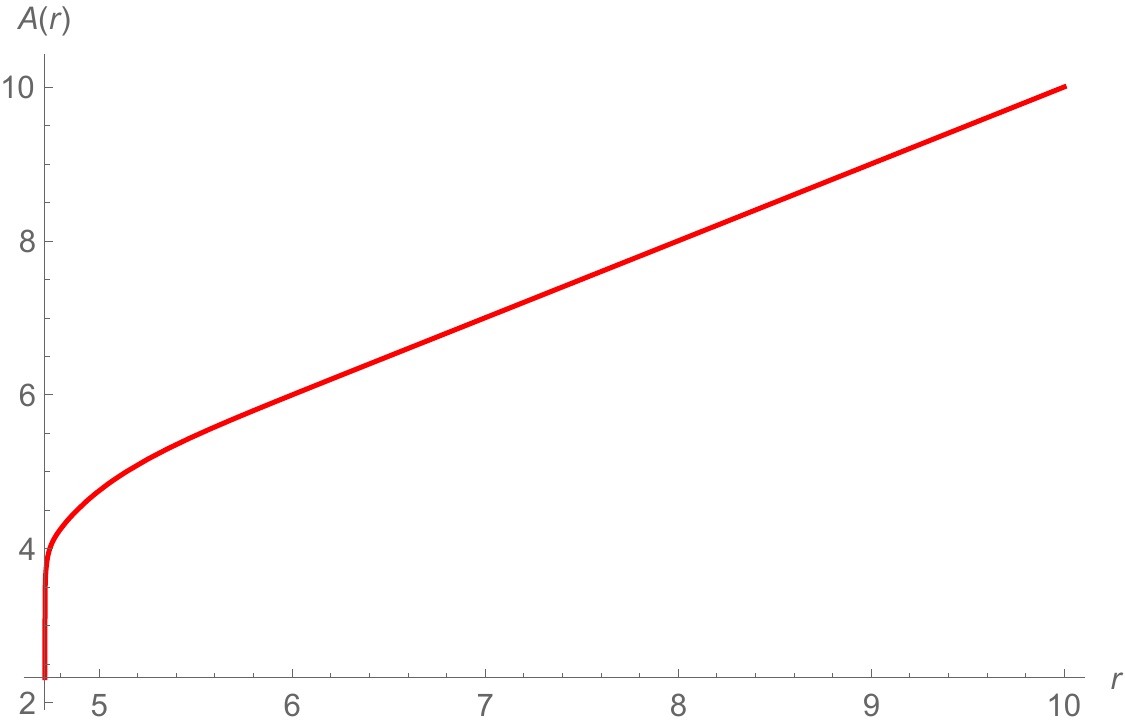}
                 \caption{Solution for $A(r)$}
         \end{subfigure}
          \begin{subfigure}[b]{0.45\textwidth}
                 \includegraphics[width=\textwidth]{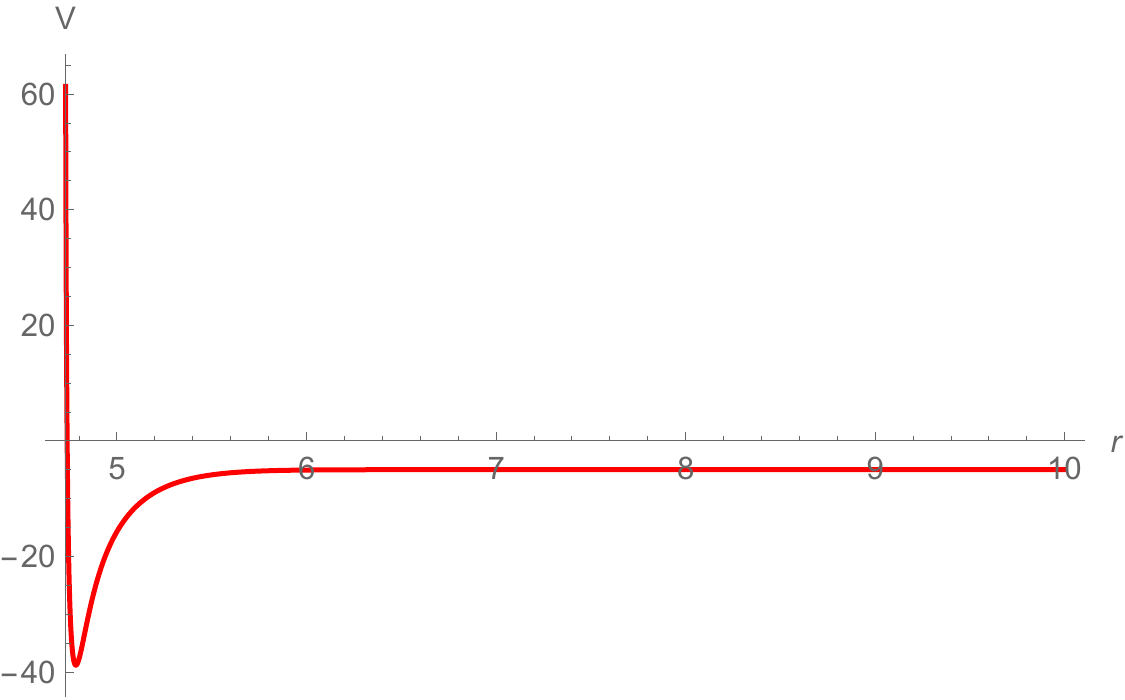}
                 \caption{Solution for $V$}
         \end{subfigure}
\caption{A holographic RG flow from supersymmetric $AdS_6$ vacuum to a non-conformal phase with $SO(2)$ symmetry in the IR for $\phi_5=0$.}\label{fig1}
 \end{figure}

In this case, we find the scalar potential
\begin{eqnarray}
V&=&\frac{1}{2}g^2e^{2(\sigma+\phi_1-\phi_2)}(e^{2\phi_1}-4e^{2\phi_2}+e^{2\phi_1+4\phi_2})+\frac{1}{2}m^2e^{-6\sigma-2\phi_4}(1+e^{4\phi_4})\nonumber \\
& &+gme^{-2\sigma-\phi_2-\phi_4}(1+e^{2\phi_4})(e^{2\phi_1+2\phi_2}-e^{2\phi_1}-2e^{2\phi_2}).
\end{eqnarray}
The BPS equations are given by
\begin{eqnarray}
& &\phi'_1=2ge^{\sigma+2\phi_1}\sinh\phi_2,\nonumber \\
& &\phi'_2=-2ge^\sigma(e^{\phi_2}-e^{2\phi_1}\cosh\phi_2),\nonumber \\
& &\phi'_4=-2me^{-3\sigma}\sinh\phi_4,\nonumber \\
& &\sigma'=-\frac{1}{2}ge^\sigma\left[\cosh\phi_2-\sinh\phi_2(e^{2\phi_1}-1)\right]+\frac{3}{2}me^{-3\sigma}\cosh\phi_4,\nonumber \\
& &A'=\frac{1}{4}ge^\sigma(e^{2\phi_1-\phi_2}+2e^{\phi_2}-e^{2\phi_1+\phi_2})+\frac{1}{2}me^{-3\sigma}\cosh\phi_4\, .
\end{eqnarray}
A truncation with $\phi_2=\phi_1=\phi$ and $\phi_4=\varphi$ has already been considered in the previous section. We now look for more general solutions with all four active scalars. However, in this case, we are not able to analytically find the flow solution. Accordingly, we will find the solution, numerically. We first look at asymptotic behaviors of scalars near the $AdS_6$ critical point. These are given by
\begin{eqnarray}
\phi_1+\phi_2\sim e^{\frac{3r}{L}},& &\qquad 2\phi_2-\phi_1\sim e^{-\frac{6r}{L}},\nonumber \\
\phi_4\sim e^{-\frac{r}{L}},& &\qquad \sigma\sim e^{-\frac{3r}{L}}\, .
\end{eqnarray}
This result is consistent with the scalar masses given in the previous section. In particular, the $SO(3)$ singlet $\phi_1+\phi_2$ corresponds to an irrelevant operator of dimension $\Delta=8$. On the other hand, $\sigma$ and $2\phi_2-\phi_1$, part of $\mathbf{5}$ of $SO(3)$, are dual respectively to operators of dimensions $\Delta=3$ and $\Delta=6$. These operators have non-vanishing expectation values. Finally, we also see a source term for an operator of dimension $\Delta=4$ dual to $\phi_4$.
\\  
\indent We give an example of numerical RG flow solutions in figure \ref{fig1} for $g=3m$ and $m=\frac{1}{2}$. As in the previous cases, the solutions are generally singular at a finite value of $r$. From the figure, we immediately see that the singularity is unphysical by the criterion of \cite{Gubser_singularity} since $V\rightarrow \infty$ near the singularity.

\subsubsection{Solutions with $\phi_4=0$}
\begin{figure}
         \centering
               \begin{subfigure}[b]{0.45\textwidth}
                 \includegraphics[width=\textwidth]{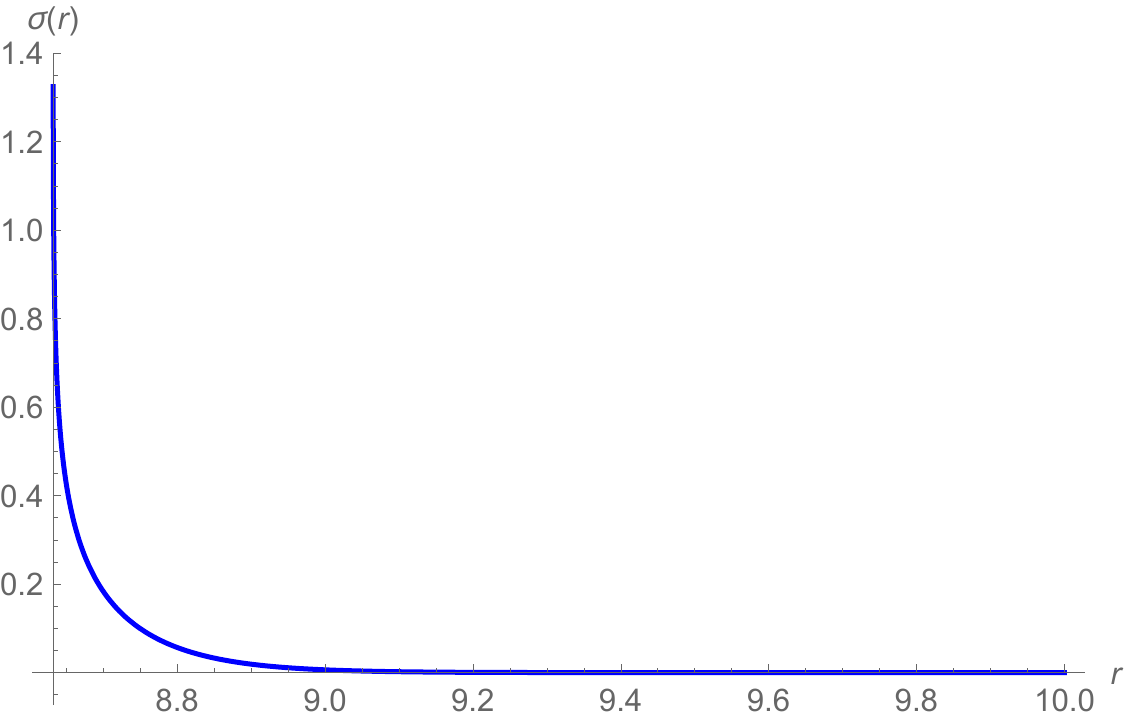}
                 \caption{Solution for $\sigma(r)$}
         \end{subfigure}
         \begin{subfigure}[b]{0.45\textwidth}
                 \includegraphics[width=\textwidth]{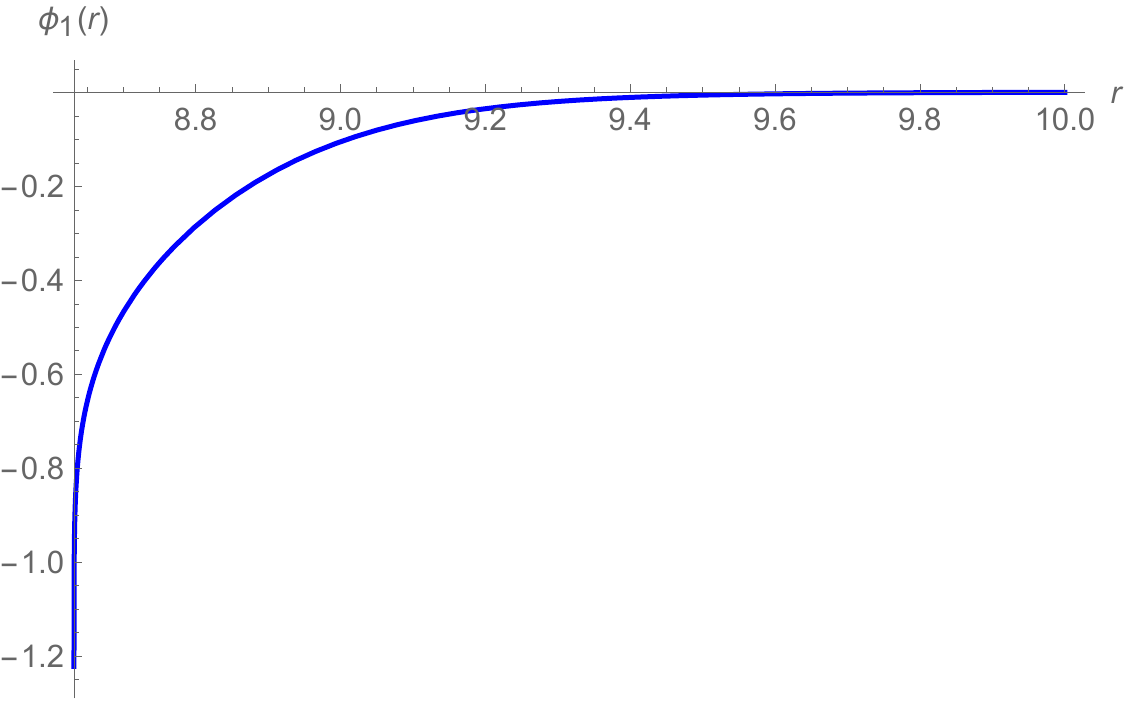}
                 \caption{Solution for $\phi_1(r)$}
         \end{subfigure}\\
          \begin{subfigure}[b]{0.45\textwidth}
                 \includegraphics[width=\textwidth]{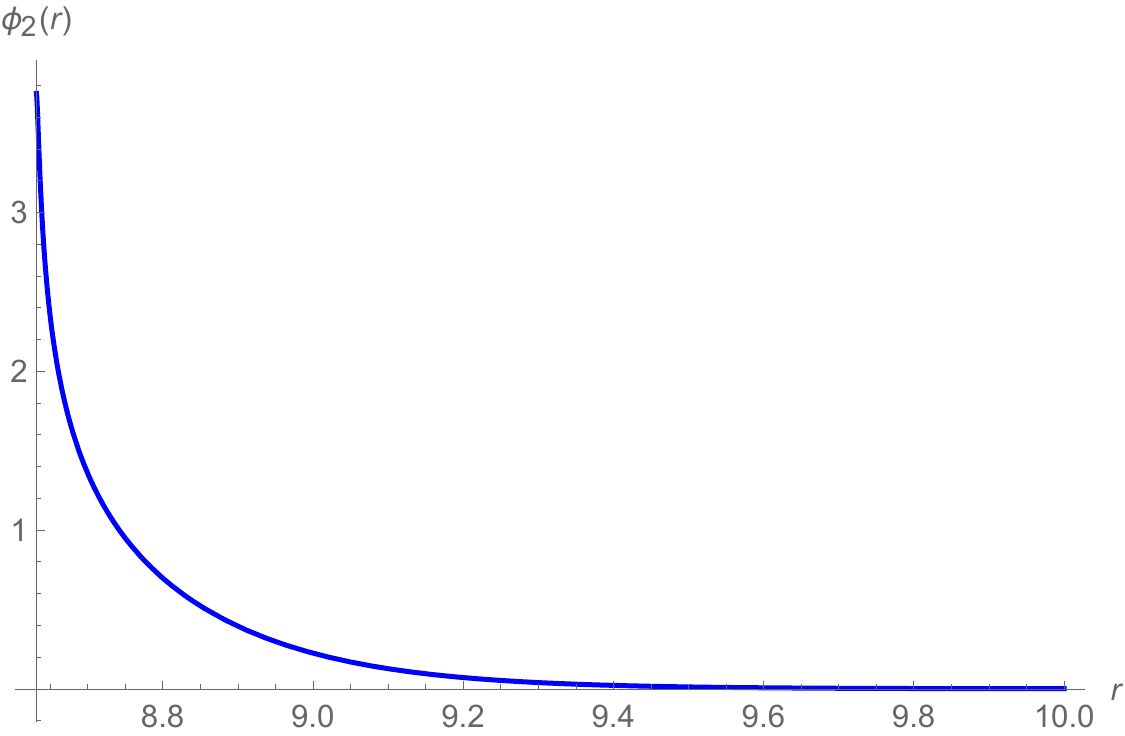}
                 \caption{Solution for $\phi_2(r)$}
         \end{subfigure}
          \begin{subfigure}[b]{0.45\textwidth}
                 \includegraphics[width=\textwidth]{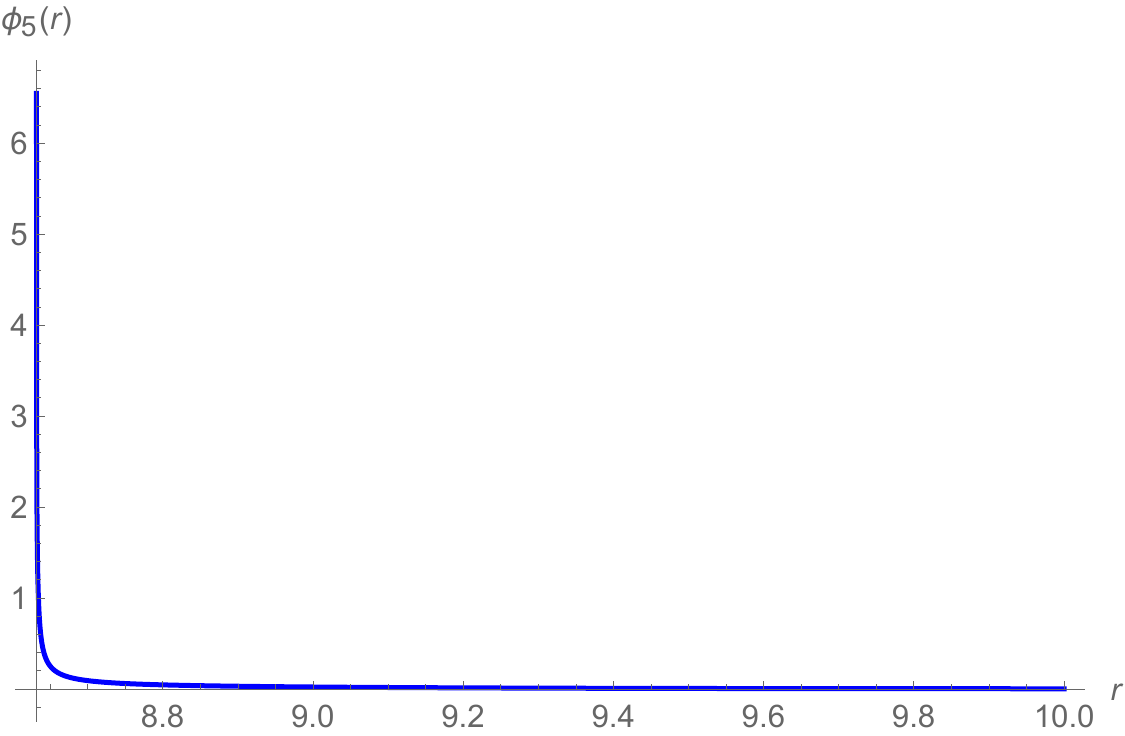}
                 \caption{Solution for $\phi_5(r)$}
         \end{subfigure}\\
         \begin{subfigure}[b]{0.45\textwidth}
                 \includegraphics[width=\textwidth]{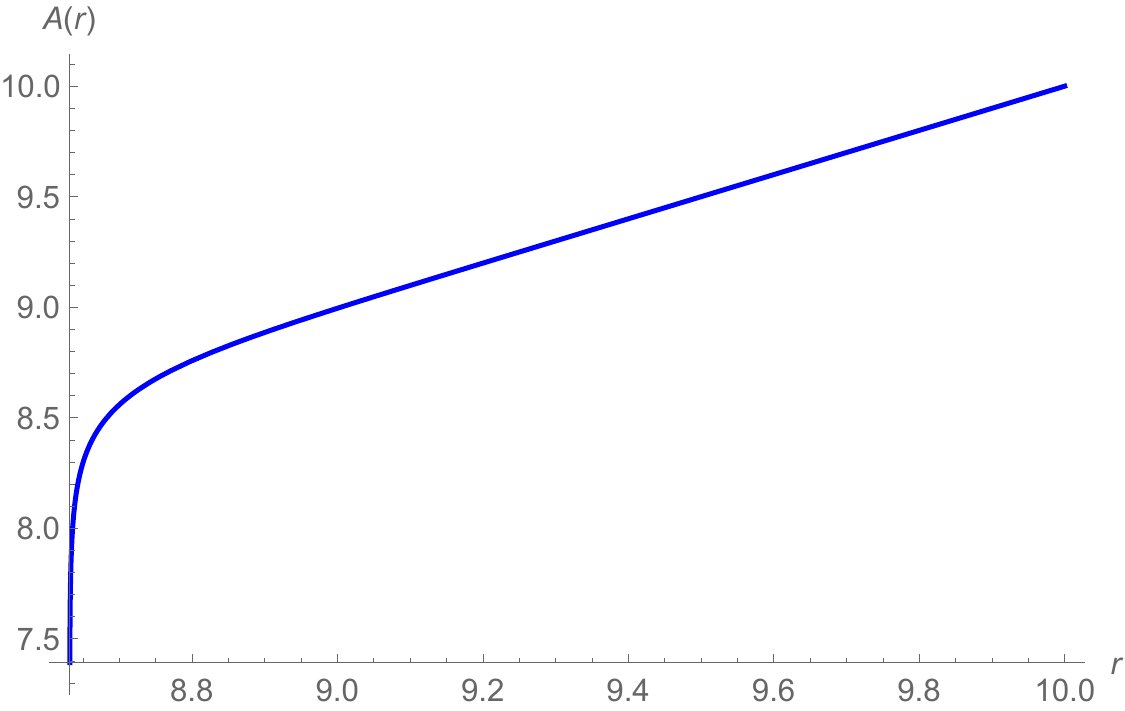}
                 \caption{Solution for $A(r)$}
         \end{subfigure}
          \begin{subfigure}[b]{0.45\textwidth}
                 \includegraphics[width=\textwidth]{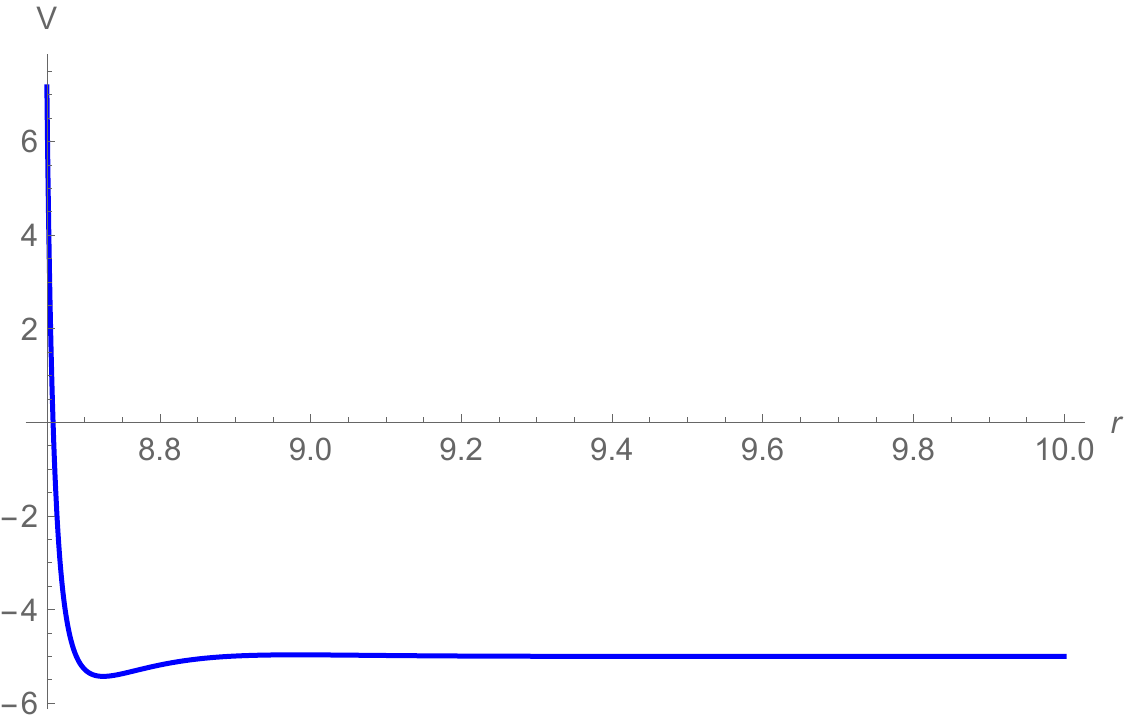}
                 \caption{Solution for $V$}
         \end{subfigure}
\caption{A holographic RG flow from supersymmetric $AdS_6$ vacuum to a non-conformal phase with $SO(2)$ symmetry in the IR for $\phi_4=0$.}\label{fig2}
 \end{figure}

Setting $\phi_4=0$, we find the scalar potential
\begin{eqnarray}
V&=&\frac{1}{8}g^2e^{2(\sigma+\phi_1-\phi_2)}\left[e^{2\phi_1}(e^{2\phi_2}-1)^2(e^{2\phi_5}+e^{-2\phi_5})+2e^{2\phi_1}(1+e^{4\phi_2})\right.\nonumber \\
& &
\left.+4e^{2\phi_2}(e^{2\phi_1}-4)\right]+gme^{-2\sigma-\phi_2-\phi_5}(1+e^{2\phi_5})\left[e^{2\phi_2}(e^{2\phi_1}-2)-e^{2\phi_1}\right]\nonumber \\
& &+m^2e^{-6\sigma}
\end{eqnarray}
and the following BPS equations
\begin{eqnarray}
& &\phi'_1=2ge^{\sigma+2\phi_1}\cosh\phi_5\sinh\phi_2,\nonumber \\
& &\phi'_2=2ge^\sigma\textrm{sech}\phi_5\left[(e^{2\phi_1}-1)\cosh\phi_2-\sinh\phi_2\right],\nonumber \\
& &\phi'_5=-2ge^\sigma\sinh\phi_5\left[\cosh\phi_2+(1-e^{2\phi_1})\sinh\phi_2\right],\nonumber \\
& &\sigma'=\frac{3}{2}me^{-3\sigma}-\frac{1}{2}ge^\sigma\cosh\phi_5(\cosh\phi_2-2e^{\phi_1}\sinh\phi_1\sinh\phi_2),\nonumber \\
& &A'=\frac{1}{2}me^{-3\sigma}+\frac{1}{2}ge^\sigma\cosh\phi_5(\cosh\phi_2-2e^{\phi_1}\sinh\phi_1\sinh\phi_2).
\end{eqnarray}
We first consider a simple solution with $\phi_1=\phi_2=0$. In this case, the solution is given by
\begin{eqnarray}
& &\phi_5=\ln\left[\frac{1+e^{-2g(\rho-\rho_0)}}{1-e^{-2g(\rho-\rho_0)}}\right],\nonumber \\
& & \sigma=\frac{1}{4}\ln\left[\frac{(\sigma_0+3m\coth\phi_5)\sinh\phi_5}{g}\right],\nonumber \\
& &A=\frac{1}{12}\ln\left[3m\cosh\phi_5+\sigma_0\sinh\phi_5\right]-\frac{1}{3}\ln\sinh\phi_5
\end{eqnarray}
in which $\sigma_0$ and $\rho_0$ are integration constants. The coordinate $\rho$ is defined by $\frac{d\rho}{dr}=e^\sigma$. At $\rho\rightarrow \infty$, we find
\begin{equation}
\phi_5\sim e^{-6mr}\sim e^{-\frac{3r}{L}},\qquad \sigma \sim \frac{\sigma_0}{3m}e^{-\frac{3r}{L}},\qquad A\sim 2mr
\end{equation}
in which we have set $g=3m$. From these behaviors, we see that $\sigma$ and $\phi_5$ are both dual to operators of dimension $\Delta=3$, but there are no source terms for these operators turned on. The flow is then driven by vacuum expectation values of these two operators. 
\\
\indent As $\rho\rightarrow \rho_0$, the solution is singular with
\begin{eqnarray}
\phi_5&\sim& -\ln[6m(\rho-\rho_0)],\nonumber \\ 
\sigma&\sim& \frac{1}{4}\ln\left[\left(\frac{\sigma_0+3m}{36m^2}\right)(\rho-\rho_0)^{-1}+(3m-\sigma_0)(\rho-\rho_0)\right],\nonumber \\
A&\sim& \frac{1}{12}\ln\left[\frac{4(\sigma_0+3m)}{3m}(\rho-\rho_0)^{-1}+48m(3m-\sigma_0)(\rho-\rho_0)\right]\nonumber \\
& &+\frac{1}{3}\ln(\rho-\rho_0). 
\end{eqnarray} 
For $\sigma_0\neq -3m$, we find
\begin{equation}
\sigma\sim -\frac{1}{4}\ln(\rho-\rho_0),\qquad A\sim \frac{1}{4}\ln(\rho-\rho_0)
\end{equation}
while for $\sigma_0=-3m$, the solution becomes
\begin{equation}
\sigma\sim \frac{1}{4}\ln(\rho-\rho_0),\qquad A\sim \frac{5}{12}\ln(\rho-\rho_0).
\end{equation}
Both of these give $V\rightarrow -\infty$, so these are physically acceptable singularities and should be dual to non-conformal phases of the dual $N=2$ SCFT in five dimensions.
\\
\indent For more general solutions with all scalars non-vanishing, we need to numerically find the solutions. The asymptotic behaviors of scalar fields near the $AdS_6$ critical point can be found as in the previous case. The result is given by
\begin{eqnarray} 
& &\phi_1+\phi_2\sim e^{\frac{3r}{L}},\qquad 2\phi_2-\phi_1\sim e^{-\frac{6r}{L}},\nonumber \\
& &\phi_5\sim e^{-\frac{3r}{L}},\qquad \sigma\sim e^{-\frac{3r}{L}}\, .
\end{eqnarray}
The only difference from the previous case is the fact that the source term for a dimension $4$ operator is replaced by a vacuum expectation value of a dimension $3$ operator dual to $\phi_5$. With numerical values $g=3m$ and $m=\frac{1}{2}$ as in the previous case, we find an example of numerical RG flow solutions as shown in figure \ref{fig2}. Again, from the figure, we see that the singularity is unphysical by the criterion of \cite{Gubser_singularity}.
  
\section{Supersymmetric Janus solutions}\label{Janus}
In this section, we consider supersymmetric Janus solutions from the matter-coupled $F(4)$ gauged supergravity with $ISO(3)\times U(1)$ gauge group. These solutions holographically describe four-dimensional conformal interfaces within the five-dimensional $N=2$ SCFT dual to the supersymmetric $AdS_6$ vacuum. The metric ansatz takes the form of an $AdS_5$-sliced domain wall given by
\begin{equation}
ds^2=e^{2f(r)}d\tilde{s}^2+dr^2\label{Janus_ansatz}
\end{equation}  
with $d\tilde{s}^2$ being the metric on a unit $AdS_5$ space. It turns out that the $SO(3)$ singlet scalars do not lead to Janus solutions. We then consider the $SO(2)$ singlet sector with the coset representative given in \eqref{L_SO2}. As in the case of holographic RG flows, we need to set $\phi_3=0$ in order to consistently truncate out all the vector fields. 
\\
\indent We now solve the BPS conditions arising from setting the supersymmetry transformations of fermionic fields $(\psi^A_\mu,\chi^A,\lambda^{IA})$ to zero. We begin with the gaugino variations, $\delta\lambda^{IA}=0$. For $I=1,2,3,4$, these variations lead to the following conditions 
\begin{eqnarray}
\delta\lambda^{1A}&:&\quad \phi_1'{(\sigma_1)^A}_B\gamma_{\hat{r}}\epsilon^B+M_1{(\sigma_1)^A}_B\epsilon^B+i\gamma_7M_2{(\sigma_2)^A}_B\epsilon^B=0,\label{dLambda1}\\
\delta\lambda^{2A}&:&\quad \phi_1'{(\sigma_2)^A}_B\gamma_{\hat{r}}\epsilon^B-i\gamma_7M_2{(\sigma_1)^A}_B\epsilon^B+M_1{(\sigma_2)^A}_B\epsilon^B=0,\label{dLambda2}\\
\delta\lambda^{3A}&:&\quad \left[(\cosh\phi_2\sinh\phi_4\sinh\phi_5\phi_0'+\cosh\phi_5\phi_2'){(\sigma_3)^A}_B\right.\nonumber \\
& &\quad \left.-\cosh\phi_2\cosh\phi_4\phi'_0\delta^A_B\gamma_7\right]\gamma_{\hat{r}}\epsilon^B+M_0\gamma_7\epsilon^A+M_3{(\sigma_3)^A}_B\epsilon^B=0,
\\
\delta\lambda^{4A}&:&\quad \left[-(\cosh\phi_4\sinh\phi_2\sinh\phi_5\phi'_0+\cosh\phi_5\phi_4')\delta^A_B\gamma_7\right.\nonumber \\
& &\quad\left.+(\phi_5'-\sinh\phi_2\sinh\phi_4\phi_0'){(\sigma_3)^A}_B\right]\gamma_{\hat{r}}\epsilon^B +\widetilde{M}_0\gamma_7\epsilon^A+\widetilde{M}_3{(\sigma_3)^A}_B\epsilon^B=0\nonumber \\
& &\label{dLambda3}
\end{eqnarray}
with
\begin{eqnarray}
& &M_0=2me^{-3\sigma}\sinh\phi_0\cosh\phi_2,\nonumber \\
& &\widetilde{M}_0=2me^{-3\sigma}(\cosh\phi_0\cosh\phi_5\sinh\phi_4+\sinh\phi_0\sinh\phi_2\sinh\phi_5),\nonumber \\ 
& &M_1=2ge^{\sigma+2\phi_1}(\cosh\phi_0\cosh\phi_5\sinh\phi_2+\sinh\phi_0\sinh\phi_4\sinh\phi_5),\nonumber \\
& &M_2=-2ge^{\sigma+2\phi_1}\sinh\phi_0\cosh\phi_4,\nonumber \\
& &M_3=2ge^\sigma\left[(e^{2\phi_1}-1)\cosh\phi_0\cosh\phi_2-\sinh\phi_2\right],\nonumber \\
& &\widetilde{M}_3= 4ge^{\sigma+\phi_1}\sinh\phi_1\left(\cosh\phi_5\sinh\phi_0\sinh\phi_4+\cosh\phi_0\sinh\phi_2\sinh\phi_5\right)\nonumber \\
& &\qquad\,\,\,-2ge^\sigma\cosh\phi_2\sinh\phi_5\, .\label{M_def}
\end{eqnarray}
In addition, $\delta \chi^A=0$ conditions and the gravitino variations lead to
\begin{eqnarray}
-\frac{1}{2}\sigma'\gamma_{\hat{r}}\epsilon^A&=&N_0\epsilon^A+N_3\gamma_7{(\sigma_3)^A}_B\epsilon^B,\\ \label{dchi_eq}
 D_\mu\epsilon^A&=&S_0\gamma_\mu \epsilon^A+S_3{(\sigma_3)^A}_B\gamma_7\gamma_\mu \epsilon^B
\end{eqnarray}
with
\begin{eqnarray}
N_0&=&\frac{3}{4}me^{-3\sigma}\cosh\phi_0\cosh\phi_4-\frac{1}{4}ge^\sigma\cosh\phi_2\cosh\phi_5\nonumber \\ 
& &+\frac{1}{2}ge^{\sigma+\phi_1}\sinh\phi_1(\cosh\phi_0\cosh\phi_5\sinh\phi_2+\sinh\phi_0\sinh\phi_4\sinh\phi_5),\nonumber \end{eqnarray}
\begin{eqnarray}
N_3&=&-\frac{3}{4}me^{-3\sigma}(\cosh\phi_4\sinh\phi_0\sinh\phi_2+\cosh\phi_0\sinh\phi_4\sinh\phi_5)\nonumber \\
& &+\frac{1}{2}ge^{\sigma+\phi_1}\cosh\phi_4\sinh\phi_0\sinh\phi_1,\nonumber \\
S_0&=&\frac{1}{2}ge^{\sigma+\phi_1}\sinh\phi_1(\cosh\phi_0\cosh\phi_5\sinh\phi_2+\sinh\phi_0\sinh\phi_4\sinh\phi_5)\nonumber \\
& &-\frac{1}{4}me^{-3\sigma}\cosh\phi_0\cosh\phi_4-\frac{1}{4}ge^\sigma\cosh\phi_2\cosh\phi_5,\nonumber \\
S_3&=&-\frac{1}{4}me^{-3\sigma}(\cosh\phi_5\sinh\phi_0\sinh\phi_2+\cosh\phi_0\sinh\phi_4\sinh\phi_5)\nonumber \\
& &-\frac{1}{2}ge^{\sigma+\phi_1}\cosh\phi_4\sinh\phi_0\sinh\phi_1\, .\label{N_S_def}
\end{eqnarray}
\indent Apart from additional equations from $\delta\lambda^{4A}$, all of these equations take the same form as those considered recently in \cite{6D_Janus_RG}. The analysis then proceeds along the same line as in \cite{6D_Janus_RG}, so we will mainly give the result and refer to \cite{6D_Janus_RG} for more detail. We first note that both equations \eqref{dLambda1} and \eqref{dLambda2} are the same after multiplied by suitable Pauli matrices. The resulting equation leads to a projector of the form
\begin{equation}
-\phi'_1\gamma_{\hat{r}}\epsilon^A=M_1\epsilon^A-M_2\gamma_7{(\sigma_3)^A}_B\epsilon^B\, .\label{gamma_r_proj}
\end{equation}
Consistency then requires
\begin{equation}
{\phi'_1}^2=M_1^2+M_2^2\, .\label{phi1_eq}
\end{equation}
By using the projector \eqref{gamma_r_proj} in the remaining equations from $\delta\lambda^I_A$, we find
\begin{eqnarray}
\cosh\phi_2\cosh\phi_4\phi_0'\phi_1'&=&-M_0M_1-M_2M_3,\\
\phi_1'\phi_2'&=&\textrm{sech}\phi_5(M_1M_3-M_0M_2)\nonumber \\
& &+\tanh\phi_4\tanh\phi_5(M_0M_1+M_2M_3),\\
\cosh\phi_5\phi_1'\phi_4'&=&-(M_1\widetilde{M}_0+M_2\widetilde{M_3})\nonumber \\
& &+\sinh\phi_5\tanh\phi_2(M_0M_1+M_2M_3),\label{Janus_eq4}\\
\phi_1'\phi_5'&=&M_1\widetilde{M}_3-M_2\widetilde{M}_0\nonumber \\
& &-\tanh\phi_2\tanh\phi_4(M_0M_1+M_2M_3).\label{Janus_eq5}
\end{eqnarray}
\indent Similarly, the same procedure leads to the following equation for the dilaton 
\begin{equation}
\sigma'\phi'_1=2(N_0M_1-N_3M_2)\label{sigma_eq}
\end{equation}
together with an algebraic constraint 
\begin{equation}
M_1N_3=-N_0M_2\, .\label{constraint1}
\end{equation}
It can also be verified that this constraint is consistent with all the previously derived BPS equations.
\\
\indent Finally, the analysis for $\delta \psi_{A\mu}$ equations is performed by using the integrability conditions for the metric \eqref{Janus_ansatz} and that for a unit $AdS_5$ 
\begin{eqnarray}
& &\left[D_m,D_n\right]\epsilon=\frac{1}{4}R_{mnpq}\gamma^{pq}\epsilon,\\
& &\left[\widetilde{D}_m,\widetilde{D}_n\right]\epsilon=\frac{1}{4}\tilde{R}_{mnpq}\tilde{\gamma}^{pq}\epsilon
\end{eqnarray}
for 
\begin{eqnarray}
R_{mnpq}&=&-\left({f'}^2+e^{-2f}\right)(g_{mp}g_{nq}-g_{mq}g_{np}),\\
\textrm{and}\qquad \tilde{R}_{mnpq}&=&-(\tilde{g}_{mp}\tilde{g}_{nq}-\tilde{g}_{mq}\tilde{g}_{np}).
\end{eqnarray}
In these equations, $\widetilde{D}$ and $\tilde{R}_{mnpq}$ denote the covariant derivative and Riemann tensor on a unit $AdS_5$ space, respectively. The first integrability condition gives
\begin{equation}
{f'}^2+e^{-2f}=4(S_0^2+S_3^2).\label{A_eq1}
\end{equation}
Using the second integrability condition, we arrive at the following equation
\begin{equation}
f'=-\frac{2}{\phi'_1}(S_0M_1+S_3M_2).\label{A_eq2}
\end{equation}
together with an algebraic constraint
\begin{equation}
e^{-2f}=4(S_0^2+S_3^2)-\frac{4(S_0M_1+S_3M_2)^2}{M_1^2+M_2^2}\, .\label{constraint2}
\end{equation} 
In the last equation, we have used ${\phi'_1}^2$ from \eqref{phi1_eq}. It can be verified that the constraint \eqref{constraint2} is compatible with all the previously derived BPS equations. We have also checked that all the BPS equations and the two algebraic constraints \eqref{constraint1} and \eqref{constraint2} satisfy the second-ordered field equations. All of the results are very similar to the BPS equations for Janus solutions in the compact $SO(3)\times SO(3)$ gauge group studied in \cite{6D_Janus_RG}. 
\\
\indent The resulting BPS equations are highly complicated non-linear equations. Therefore, we do not expect any possibilities to find analytic Janus solutions. In the present paper, we will only consider regular Janus solutions interpolating between $AdS_6$ vacuum on both sides of the interfaces. These solutions require suitable boundary conditions at the turning points $r=r_0$ at which $f'(r_0)=0$ and $f''(r_0)>0$. In the present case, even finding numerical Janus solutions is still very complicated due to a large number of scalars involved. Accordingly, we will first look for simpler subtruncations and consider the most general case at the end of this section. 
\\
\indent From the BPS equations given in \eqref{Janus_eq4} and \eqref{Janus_eq5}, we can see that setting $\phi_4=\phi_5=0$ is a consistent truncation. Indeed, this is just a truncation to the case of $n=3$ vector multiplets with $ISO(3)$ gauge group. In this subtruncation, however, we cannot set $\phi_0=0$ because the constraint \eqref{constraint2} implies that $e^{-2f}=0$ for $\phi_4=\phi_5=\phi_0=0$ leading to flat domain walls studied in the previous section. We also see that there are no Janus solutions with $SO(3)$ symmetry since this requires $\phi_2=\phi_1$ and $\phi_0=0$. On the other hand, setting $\phi_0=\phi_1=\phi_2=0$ is also consistent and does not exclude possible Janus solutions provided that both $\phi_4$ and $\phi_5$ are non-vanishing. As in the truncation to $n=3$ vector multiplets, there do not exist Janus solutions with $SO(3)$ symmetry due to the non-vanishing scalar $\phi_5$ which is not an $SO(3)$ singlet. 
\\
\indent It should also be pointed out that all the possible solutions preserve half of the supersymmetry or eight supercharges since there is only one projector \eqref{gamma_r_proj} imposed on the Killing spinors. Finally, before giving numerical Janus solutions, we recall that $\phi_1$, $\phi_2$ and $\phi_0$ are dual to irrelevant operators of dimensions $\Delta=6,7,8$ with the behaviors near the $AdS_6$ vacuum as $r\rightarrow \infty$ given by
\begin{eqnarray}
& &\phi_0\sim e^{\frac{2r}{L}},\qquad \phi_4\sim e^{-\frac{r}{L}},\qquad \phi_5\sim e^{-\frac{3r}{L}},\nonumber \\
& &\sigma\sim e^{-\frac{3r}{L}}\qquad \phi_1+\phi_2\sim e^{\frac{3r}{L}},\qquad 2\phi_2-\phi_1\sim e^{-\frac{6r}{L}}
\end{eqnarray}
for $g=3m$ and $L=\frac{1}{2m}$. The $AdS_6$ critical point is then a repulsive fixed point along $\phi_0$ and $\phi_1+\phi_2$ directions. To obtain regular Janus solutions that approach the $AdS_6$ critical point, we need to fine-tune the boundary conditions that set the source terms of these operators to zero. This is different from the case of compact gauge groups considered in \cite{6D_Janus} and \cite{6D_Janus_RG} in which, for the $AdS_6$ critical point at the origin of the scalar manifold, all scalars are dual to relevant operators. Therefore, finding regular Janus solutions in this case is somewhat more difficult.
\subsection{Janus solutions from $\phi_0=\phi_1=\phi_2=0$ subtruncation}
\begin{figure}
         \centering
               \begin{subfigure}[b]{0.4\textwidth}
                 \includegraphics[width=\textwidth]{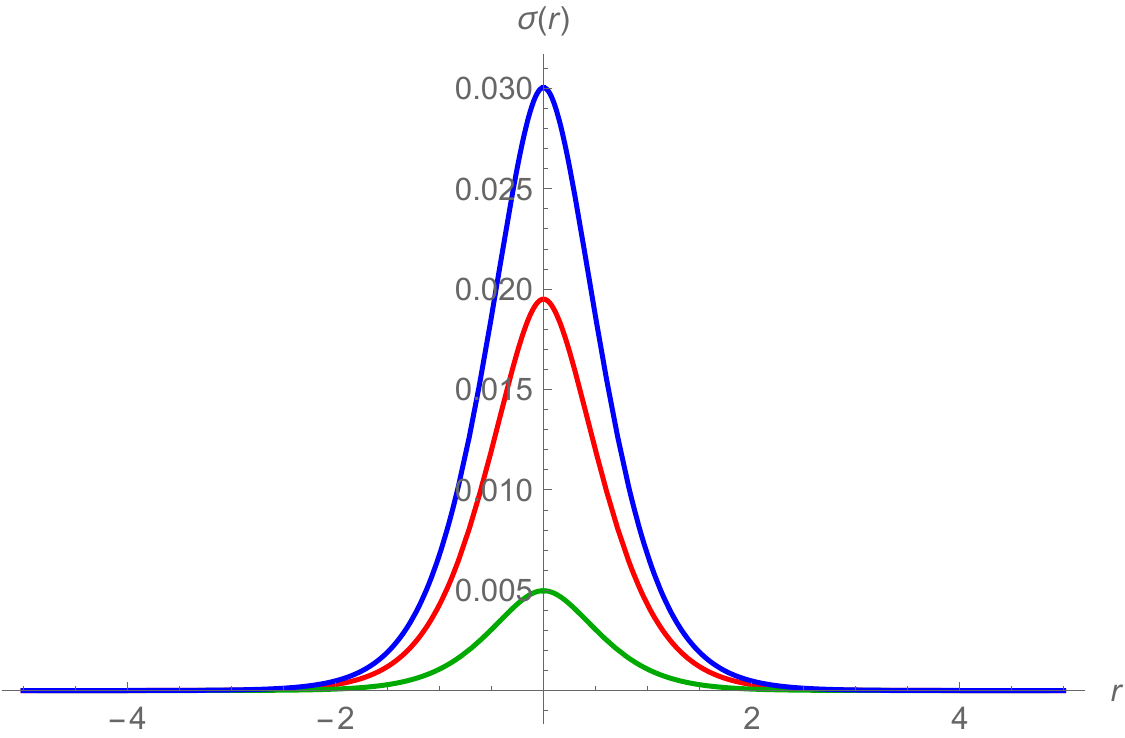}
                 \caption{Solutions for $\sigma(r)$}
         \end{subfigure}
         \begin{subfigure}[b]{0.4\textwidth}
                 \includegraphics[width=\textwidth]{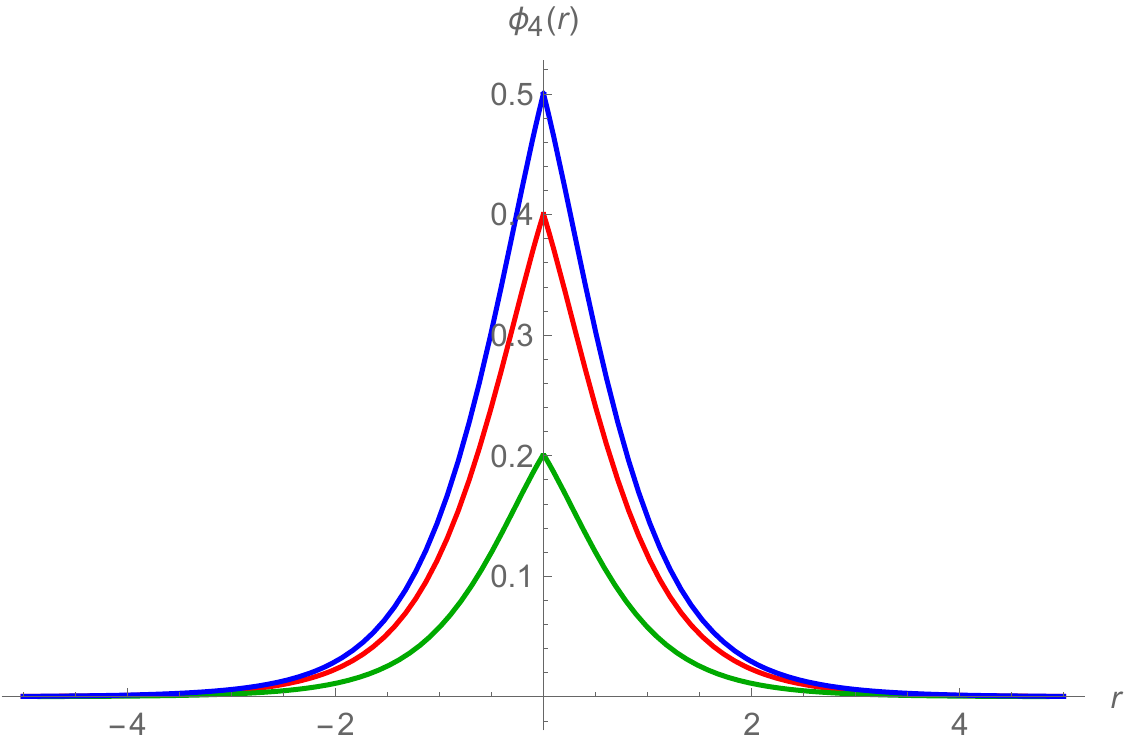}
                 \caption{Solutions for $\phi_4(r)$}
         \end{subfigure}\\
          \begin{subfigure}[b]{0.4\textwidth}
                 \includegraphics[width=\textwidth]{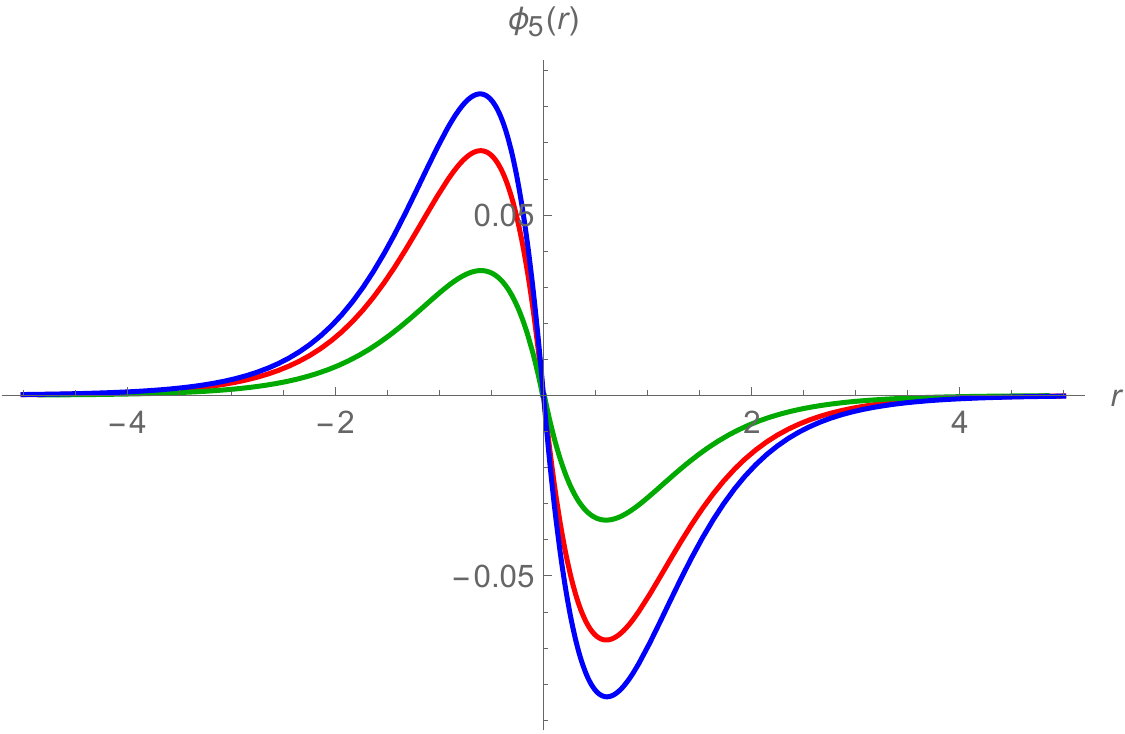}
                 \caption{Solutions for $\phi_5(r)$}
         \end{subfigure}
         \begin{subfigure}[b]{0.4\textwidth}
                 \includegraphics[width=\textwidth]{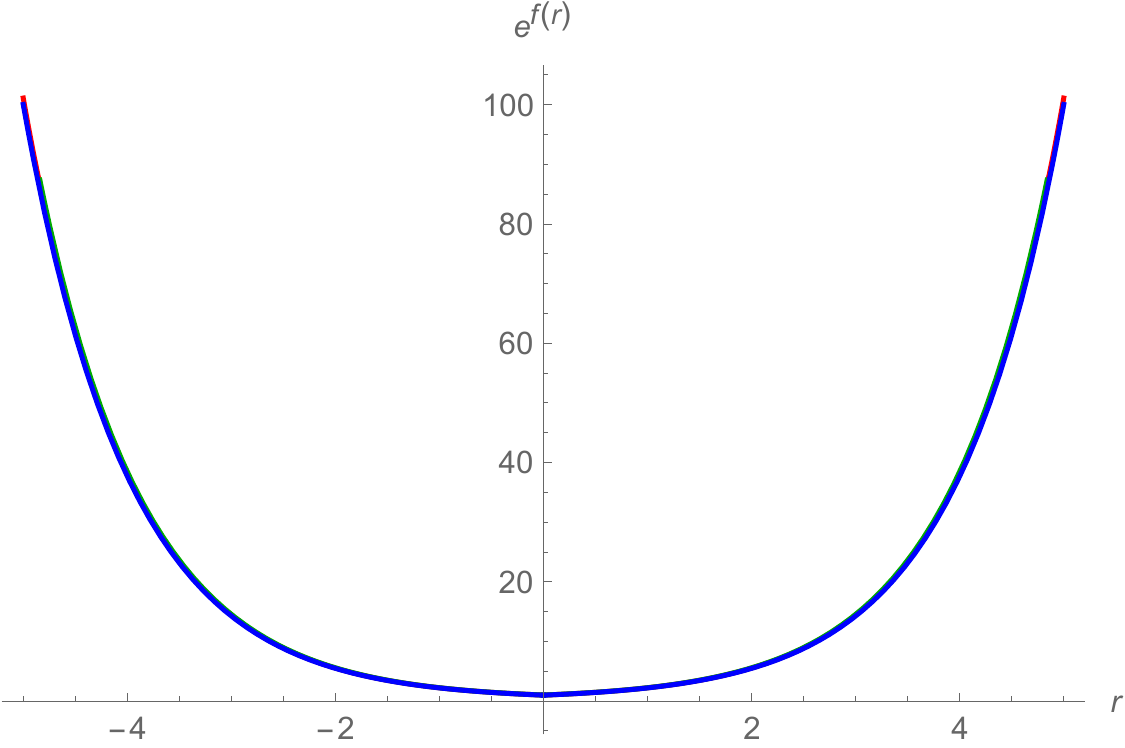}
                 \caption{Solutions for $e^{f(r)}$}
         \end{subfigure}
\caption{Supersymmetric Janus solutions preserving eight supercharges and $SO(2)\times U(1)$ symmetry from matter-coupled $F(4)$ gauged supergravity with $ISO(3)\times U(1)$ gauge group for $\phi_0=\phi_1=\phi_2=0$, $g=3m$ and $m=\frac{1}{2}$.}\label{fig3}
 \end{figure}

We begin with a simple subtruncation for $\phi_0=\phi_1=\phi_2=0$. In this case, the BPS conditions simplify considerably to
\begin{eqnarray}
\left[-\phi_5'{(\sigma_3)^A}_B+\cosh\phi_5\phi_4'\delta^A_B\gamma_7\right]\gamma_{\hat{r}}\epsilon^B &=&\widetilde{M}_0\gamma_7\epsilon^A+\widetilde{M}_3{(\sigma_3)^A}_B\epsilon^B,\nonumber \\
-\frac{1}{2}\sigma'\gamma_{\hat{r}}\epsilon^A&=&N_0\epsilon^A+N_3\gamma_7{(\sigma_3)^A}_B\epsilon^B,\nonumber \\
D_\mu\epsilon^A&=&S_0\gamma_\mu \epsilon^A+S_3{(\sigma_3)^A}_B\gamma_7\gamma_\mu \epsilon^B
\end{eqnarray}
with
\begin{eqnarray}
& &S_0=-\frac{1}{4}ge^\sigma \cosh\phi_5-\frac{1}{4}me^{-3\sigma}\cosh\phi_4,\qquad S_3=-\frac{1}{4}e^{-3\sigma}\sinh\phi_4\sinh\phi_5,\nonumber \\
& &N_0=\frac{3}{4}me^{-3\sigma}\cosh\phi_4-\frac{1}{4}ge^\sigma\cosh\phi_5,\qquad N_3=-\frac{3}{m}e^{-3\sigma}\sinh\phi_4\sinh\phi_5,\nonumber \\
& &\widetilde{M}_0=2me^{-3\sigma}\cosh\phi_5\sinh\phi_4,\qquad \widetilde{M}_3=-3ge^\sigma \sinh\phi_5\, .
\end{eqnarray}
Apart from a sign flip in $S_0$ and $S_3$ due to some difference in conventions, these equations take the same form as those considered in \cite{6D_Janus} up to a redefinition of scalars $\phi_4$ and $\phi_5$. By the same procedure as in the previous analysis, see also \cite{6D_Janus}, we arrive at the BPS equations
\begin{eqnarray}
\cosh\phi_5\phi'_4&=&-(G_0\widetilde{M}_0+G_3\widetilde{M}_3),\\
\phi'_5&=&G_0\widetilde{M_3}-G_3\widetilde{M}_0,\\
\sigma'&=&\frac{2N_0}{G_0},\\
f'&=&-2(G_0S_0+G_3S_3)
\end{eqnarray}
together with an algebraic constraint
\begin{eqnarray}
e^{-2f}=4(S_0^2+S_3^2)-4(G_0S_0+G_3S_3)^2.
\end{eqnarray}
In these equations, as in \cite{6D_Janus}, we have defined the following quantities
\begin{equation}
G_0=\eta\frac{N_0}{\sqrt{N_0^2+N_3^2}}\qquad \textrm{and}\qquad G_3=-\eta\frac{N_3}{\sqrt{N_0^2+N_3^2}}
\end{equation}
for $\eta=\pm 1$. We also note that the other algebraic constraint in \eqref{constraint1} is absent in this case. This is due to the fact that this constraint arises from the compatibility between the BPS equations from $\delta\lambda^{1,2}_A$ and $\delta\chi_A$, and in the present subtruncation, $\delta\lambda^{1,2}_A$ conditions vanish identically. 
\\
\indent We now give examples of numerical solutions for Janus interfaces. We will follow the usual procedure for finding Janus solutions by first choose suitable values of the warp factor $f(r)$ and scalar fields at the turning point $r=r_0$. At this turning point, we require $f'(r_0)=0$ and $f''(r_0)>0$ in order to make the solution approach $AdS_6$ vacuum at both $r\rightarrow \infty$ and $r\rightarrow-\infty$ limits. In addition, we can also set $r_0=0$ by shifting the radial coordinate. The numerical solutions can be found by the same procedure as in \cite{6D_Janus} to which we refer for more details on the analysis of smoothness. With the numerical values of $g=3m$ and $m=\frac{1}{2}$ which effectively set the $AdS_6$ radius to one, we find examples of Janus solutions as shown in figure \ref{fig3} for three different sets of the values of scalars at the turning point. These solutions are very similar to that given in \cite{6D_Janus}. We also note that the solutions for $f(r)$ are very close to each other for all of the three choices of the boundary conditions. These Janus solutions should be dual to conformal interfaces within the five-dimensional $N=2$ SCFT arising from turning on the source term of a dimension $4$ operator dual to $\phi_4$ and position-dependent expectation values of dimension $3$ operators dual to $\sigma$ and $\phi_5$. 

\subsection{Janus solutions from $ISO(3)$ gauge group} 
\begin{figure}
         \centering
               \begin{subfigure}[b]{0.45\textwidth}
                 \includegraphics[width=\textwidth]{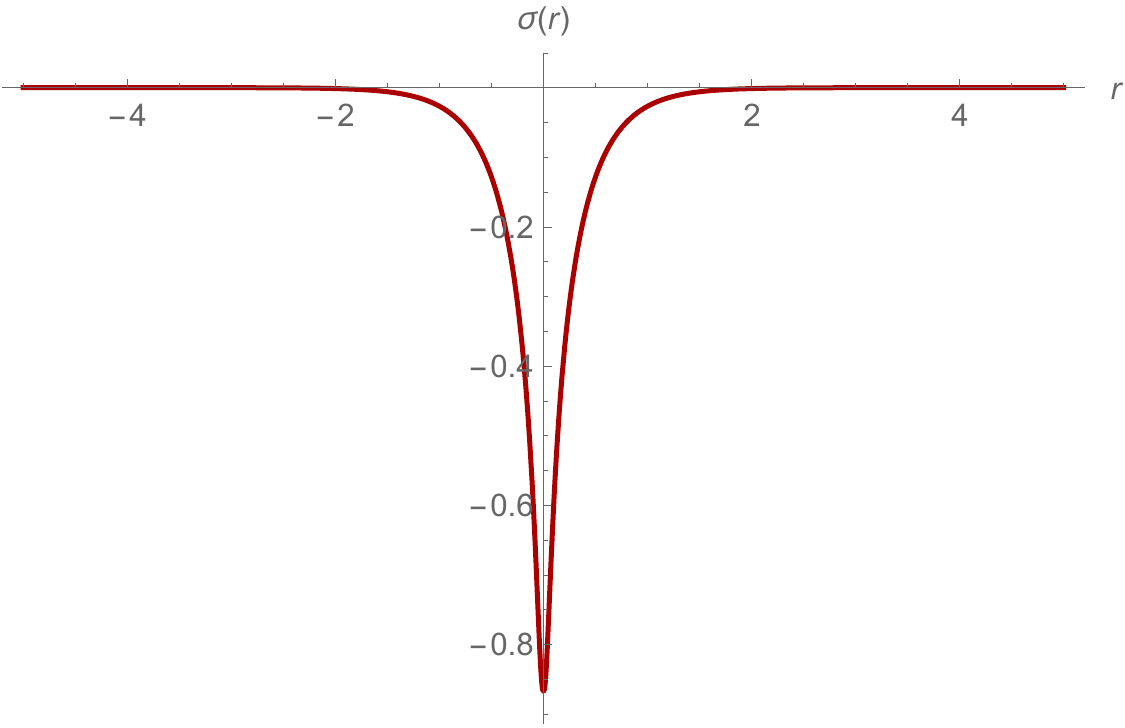}
                 \caption{Solution for $\sigma(r)$}
         \end{subfigure}
         \begin{subfigure}[b]{0.45\textwidth}
                 \includegraphics[width=\textwidth]{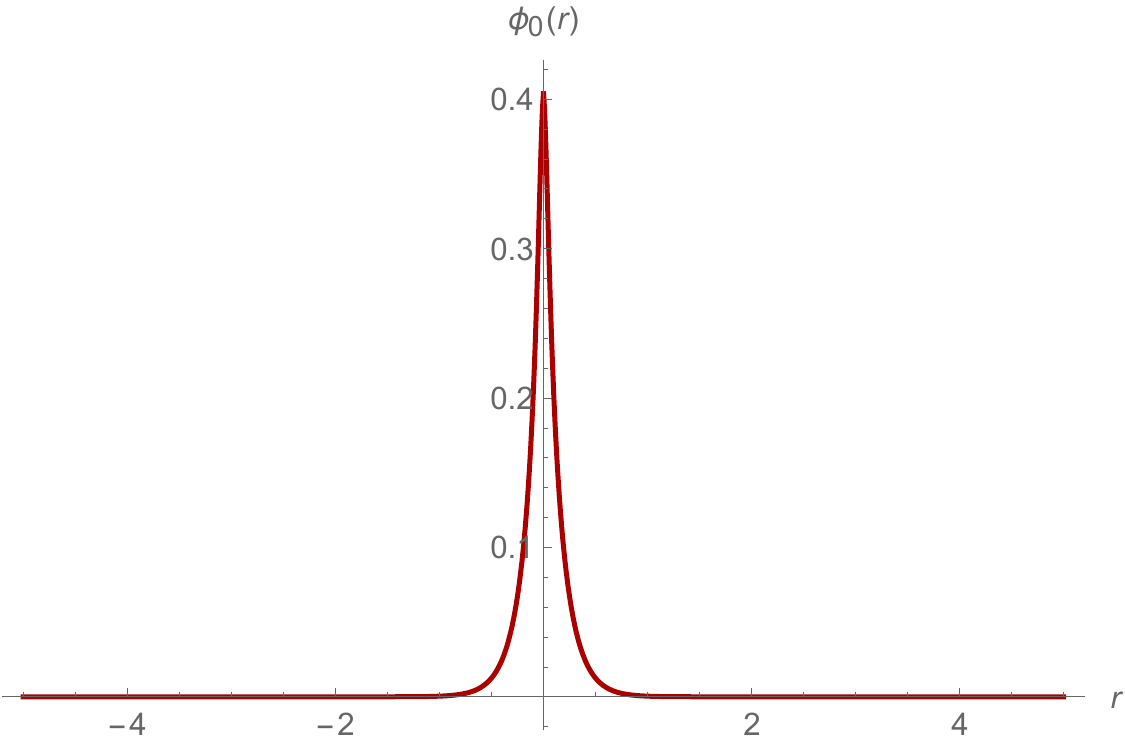}
                 \caption{Solution for $\phi_0(r)$}
         \end{subfigure}\\
          \begin{subfigure}[b]{0.45\textwidth}
                 \includegraphics[width=\textwidth]{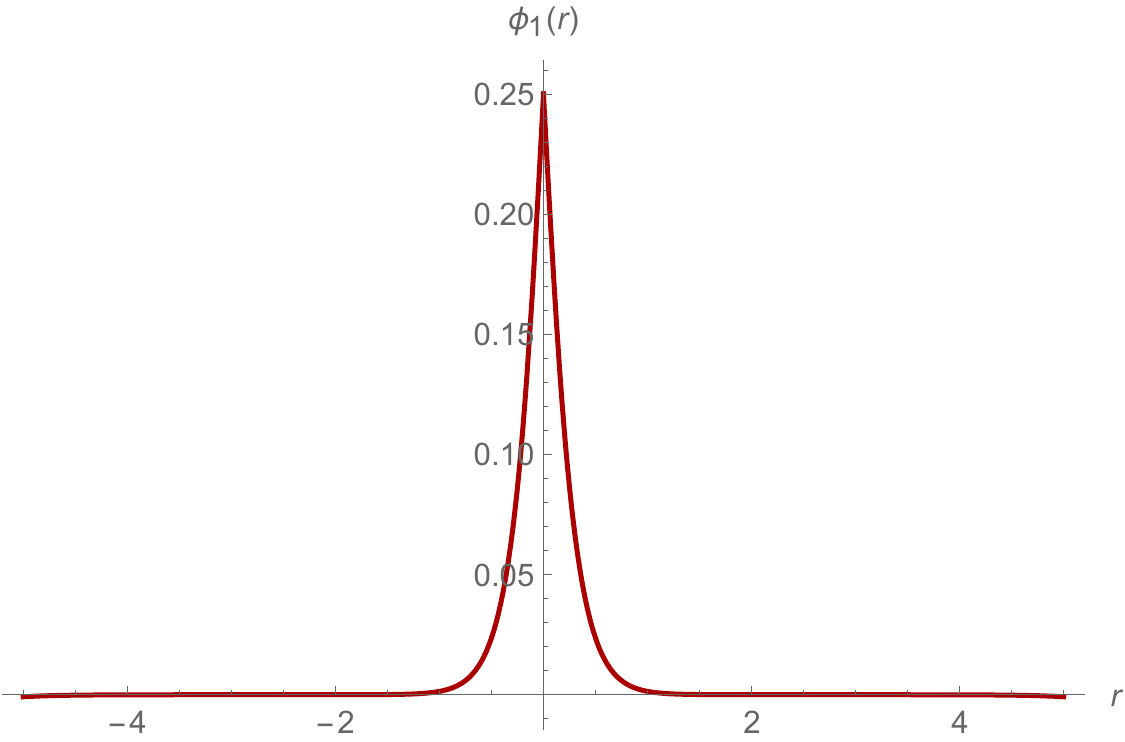}
                 \caption{Solution for $\phi_1(r)$}
         \end{subfigure}
          \begin{subfigure}[b]{0.45\textwidth}
                 \includegraphics[width=\textwidth]{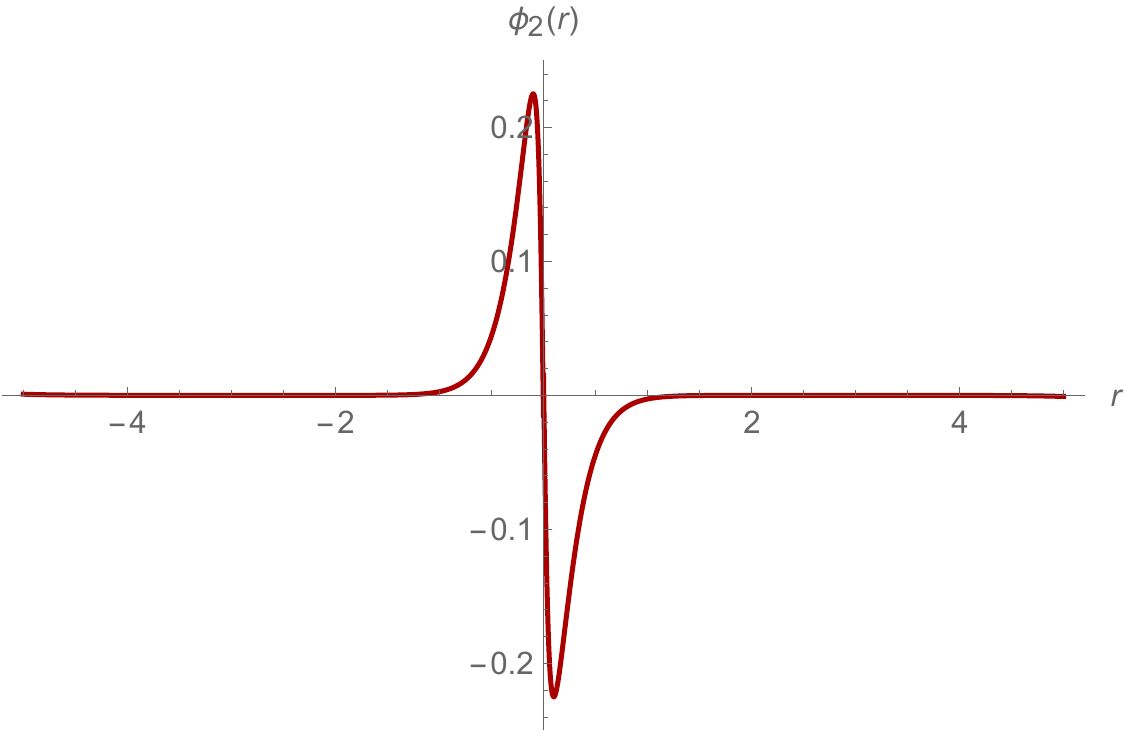}
                 \caption{Solution for $\phi_2(r)$}
         \end{subfigure}\\
         \begin{subfigure}[b]{0.45\textwidth}
                 \includegraphics[width=\textwidth]{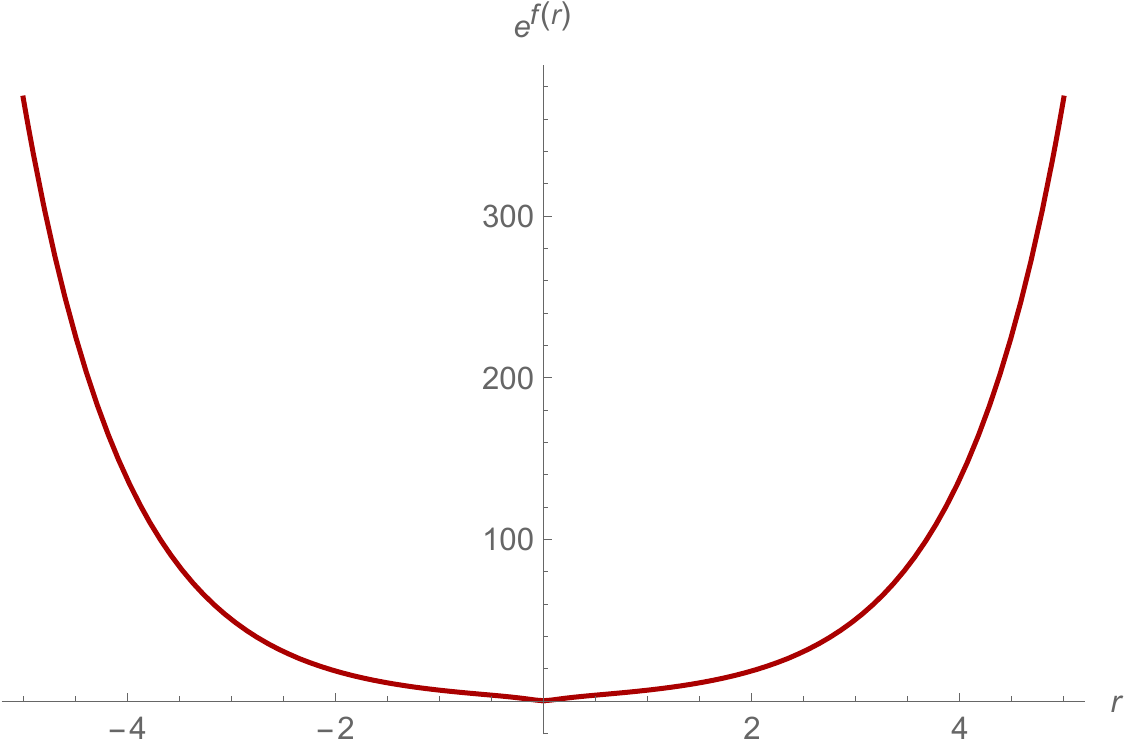}
                 \caption{Solution for $e^{f(r)}$}
         \end{subfigure}
\caption{A supersymmetric Janus solution preserving eight supercharges and $SO(2)$ symmetry from matter-coupled $F(4)$ gauged supergravity with $ISO(3)$ gauge group for $g=3m$ and $m=\frac{1}{2}$.}\label{fig4}
 \end{figure}

We now move to a more complicated truncation with $\phi_4=\phi_5=0$. The resulting solutions are effectively solutions of $F(4)$ gauged supergravity coupled to three vector multiplets with $ISO(3)$ gauge group. In this case, the BPS equations are given by
\begin{eqnarray}
\phi_1'&=&\eta \sqrt{M_1^2+M_2^2},\nonumber \\
\cosh\phi_2\phi_0'&=&-\frac{\eta(M_0M_1+M_2M_3)}{\sqrt{M_1^2+M_2^2}},\nonumber \\
\phi_2'&=&\frac{\eta(M_1M_3-M_0M_2)}{\sqrt{M_1^2+M_2^2}},\nonumber \\
\sigma'&=&\frac{2\eta (N_0M_1-N_3M_2)}{\sqrt{M_1^2+M_2^2}},\nonumber \\
f'&=&-\frac{2\eta(S_0M_1+S_3M_2)}{\sqrt{M_1^2+M_2^2}},
\end{eqnarray}
for $\eta=\pm1$, together with two algebraic constraints  
\begin{eqnarray}    
M_1N_3&=&-N_0M_2 \nonumber \\ 
\textrm{and}\qquad e^{-2f}&=&4(S_0^2+S_3^2)-4(G_0S_0+G_3S_3)^2.
\end{eqnarray}
All the quantities appearing in these equations are given in \eqref{M_def} and \eqref{N_S_def} after setting $\phi_4=\phi_5=0$. In particular, in this case, we have $\widetilde{M}_0=\widetilde{M_3}=0$.
\\
\indent It should be pointed out that the above BPS equations take a similar form to those of the compact $SO(3)\times SO(3)$ gauge group studied in \cite{6D_Janus_RG}. However, the dependence on the scalar $\phi_1$ and $\phi_2$ are different. Moreover, these scalars together with $\phi_0$ are dual to irrelevant operators whose source terms must vanish at the $AdS_6$ vacuum in order to make the solutions approach the critical point. From various attempts in numerical searching for regular Janus solutions, we find a representative solution given in figure \ref{fig4}. In this solution, we use the same values of $g=3m$ and $m=\frac{1}{2}$ as in the previous case. The solution holographically describes a conformal interface in the presence of source terms for irrelevant operators dual to $\phi_0$ and $\phi_1+\phi_2$ and position-dependent expectation values for irrelevant and relevant operators dual to $2\phi_2-\phi_1$ and $\sigma$, respectively. 
 
\subsection{Janus solutions from $ISO(3)\times U(1)$ gauge group} 
\begin{figure}
         \centering
               \begin{subfigure}[b]{0.45\textwidth}
                 \includegraphics[width=\textwidth]{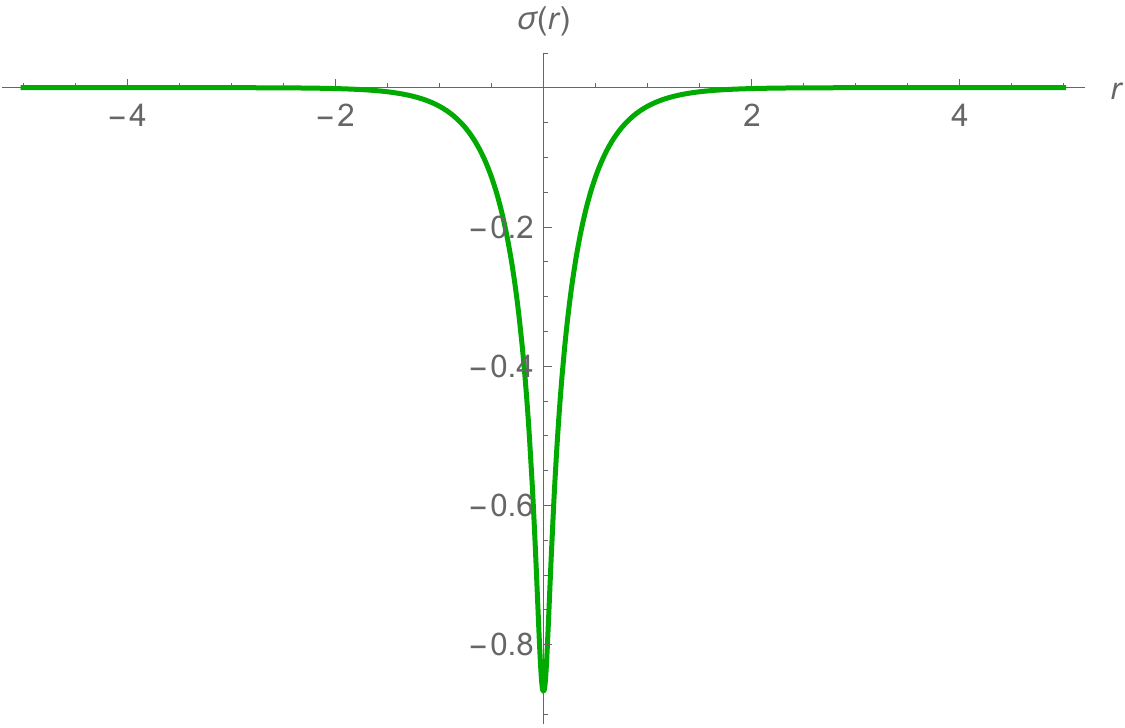}
                 \caption{Solution for $\sigma(r)$}
         \end{subfigure}
         \begin{subfigure}[b]{0.45\textwidth}
                 \includegraphics[width=\textwidth]{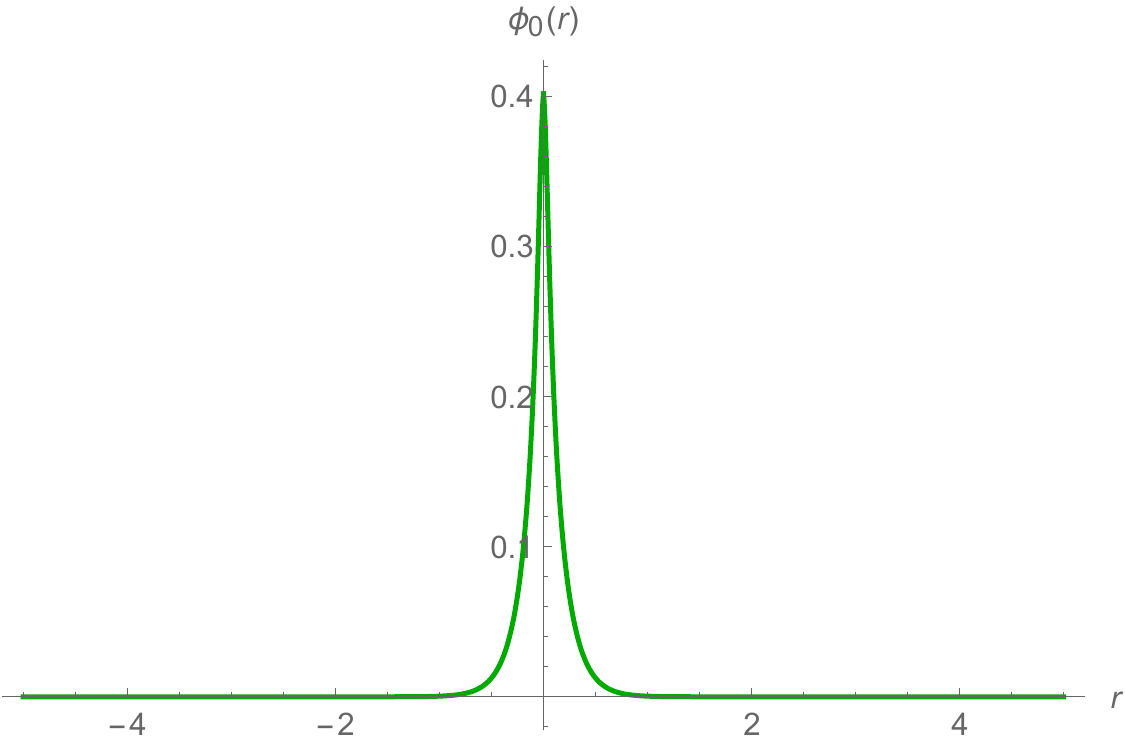}
                 \caption{Solution for $\phi_0(r)$}
         \end{subfigure}\\
          \begin{subfigure}[b]{0.45\textwidth}
                 \includegraphics[width=\textwidth]{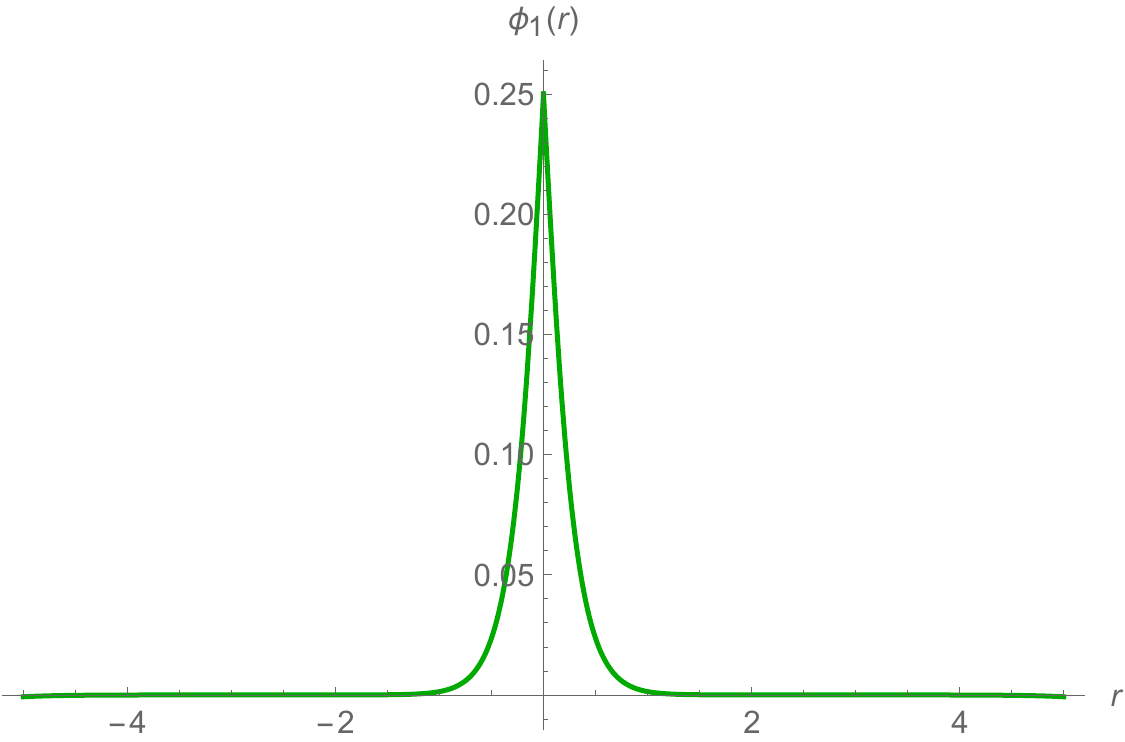}
                 \caption{Solution for $\phi_1(r)$}
         \end{subfigure}
          \begin{subfigure}[b]{0.45\textwidth}
                 \includegraphics[width=\textwidth]{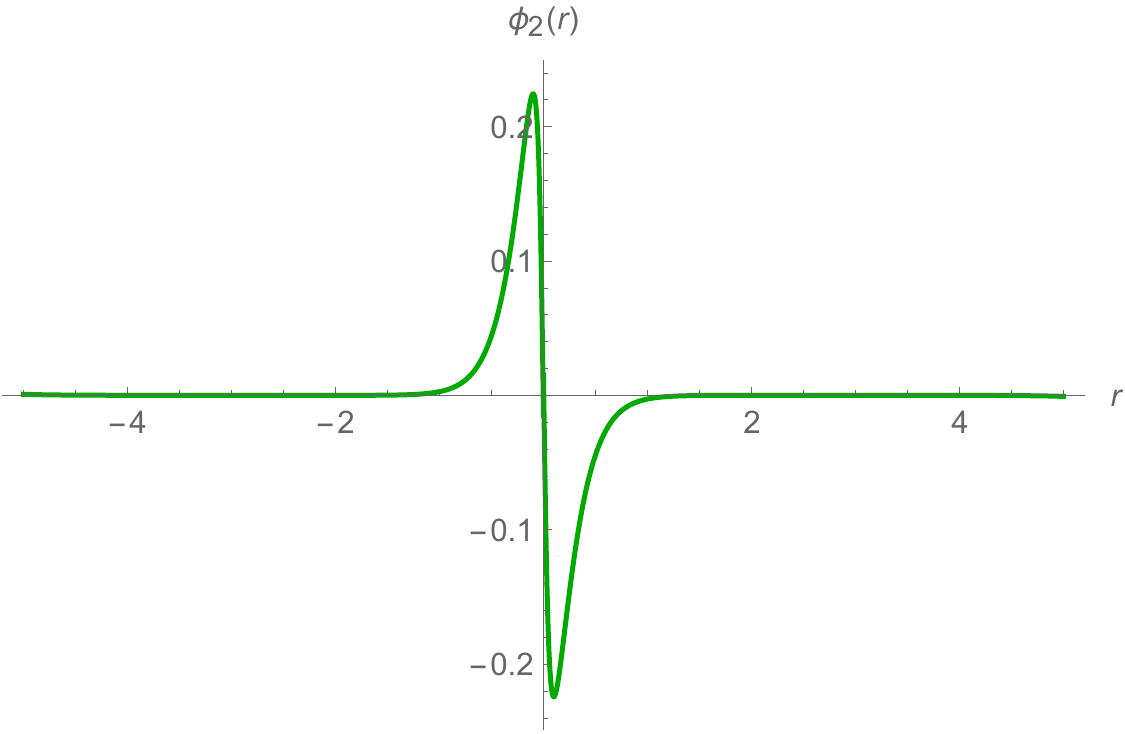}
                 \caption{Solution for $\phi_2(r)$}
            \end{subfigure}\\
          \begin{subfigure}[b]{0.45\textwidth}
                 \includegraphics[width=\textwidth]{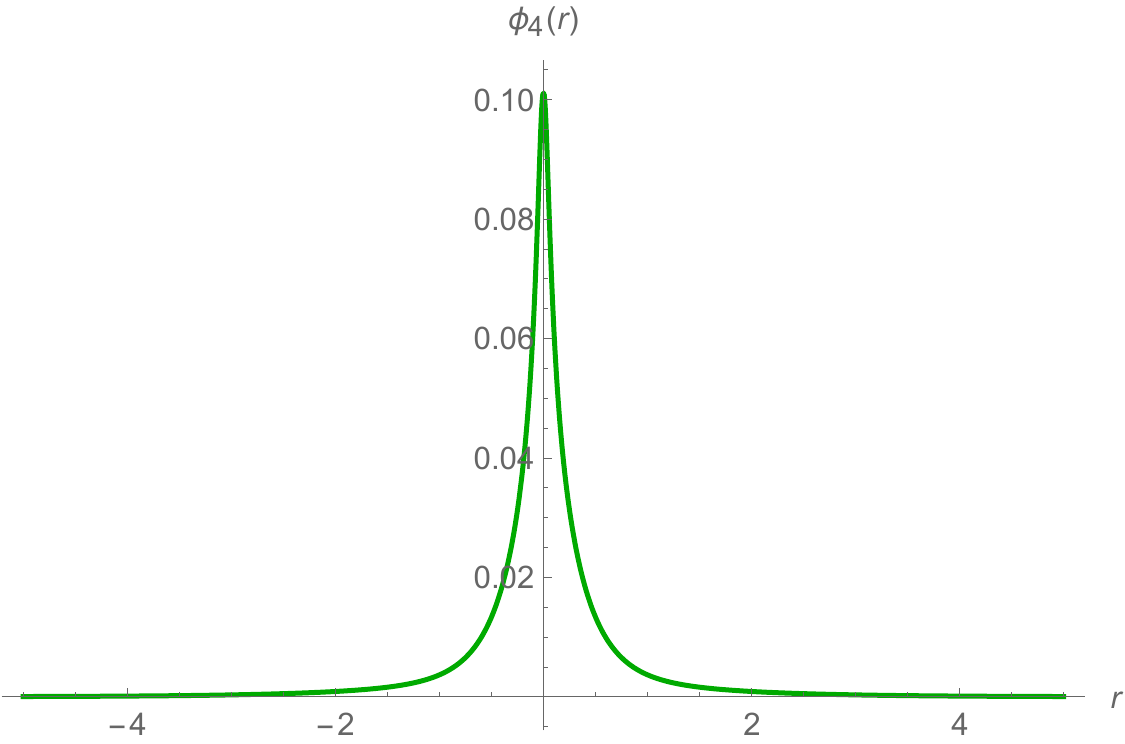}
                 \caption{Solution for $\phi_4(r)$}
         \end{subfigure}
          \begin{subfigure}[b]{0.45\textwidth}
                 \includegraphics[width=\textwidth]{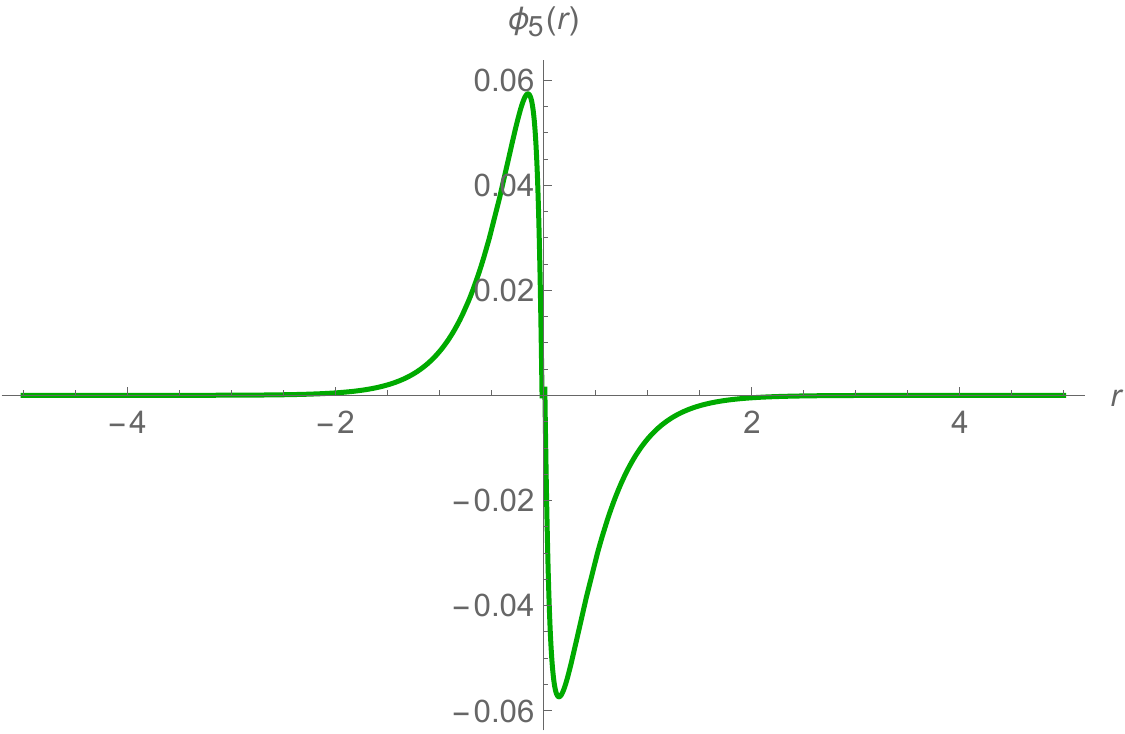}
                 \caption{Solution for $\phi_5(r)$}
         \end{subfigure}\\
         \begin{subfigure}[b]{0.45\textwidth}
                 \includegraphics[width=\textwidth]{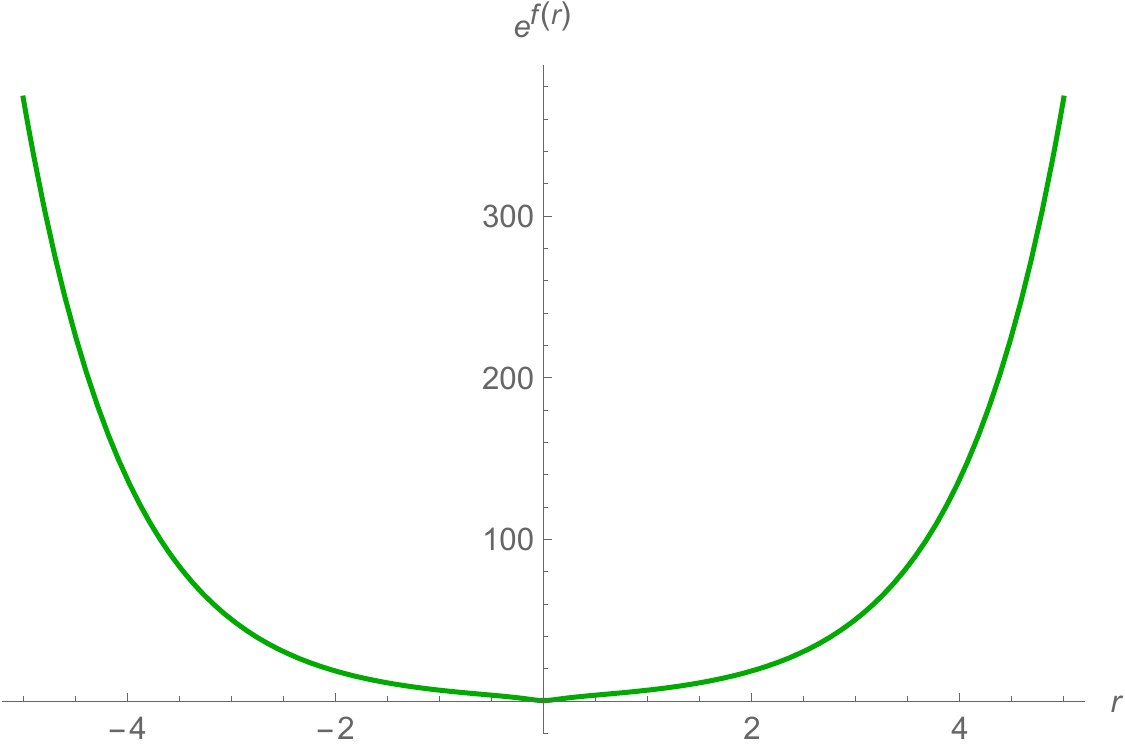}
                 \caption{Solution for $e^{f(r)}$}
         \end{subfigure}
\caption{A supersymmetric Janus solution preserving eight supercharges and $SO(2)\times U(1)$ symmetry from matter-coupled $F(4)$ gauged supergravity with $ISO(3)\times U(1)$ gauge group for $g=3m$ and $m=\frac{1}{2}$.}\label{fig5}
 \end{figure}

Finally, we consider the complete $SO(2)\times U(1)$ invariant sector of the matter-coupled $F(4)$ gauged supergravity with $ISO(3)\times U(1)$ gauge group. In this case, the BPS equations read
\begin{eqnarray}
\phi_1'&=&\eta \sqrt{M_1^2+M_2^2},\nonumber \\
\cosh\phi_2\cosh\phi_4\phi_0'&=&-\frac{\eta(M_0M_1+M_2M_3)}{\sqrt{M_1^2+M_2^2}},\nonumber\\
\phi_2'&=&\eta\frac{\textrm{sech}\phi_5(M_1M_3-M_0M_2)}{\sqrt{M_1^2+M_2^2}}\nonumber \\
& &+\eta\frac{\tanh\phi_4\tanh\phi_5(M_0M_1+M_2M_3)}{\sqrt{M_1^2+M_2^2}},\nonumber \\
\cosh\phi_5\phi_4'&=&-\eta\frac{(M_1\widetilde{M}_0+M_2\widetilde{M_3})}{\sqrt{M_1^2+M_2^2}}+\eta\frac{\sinh\phi_5\tanh\phi_2(M_0M_1+M_2M_3)}{\sqrt{M_1^2+M_2^2}},\nonumber \\
\phi_5'&=&\eta\frac{(M_1\widetilde{M}_3-M_2\widetilde{M}_0)}{\sqrt{M_1^2+M_2^2}}-\eta\frac{\tanh\phi_2\tanh\phi_4(M_0M_1+M_2M_3)}{\sqrt{M_1^2+M_2^2}},\nonumber \\
\sigma'&=&\frac{2\eta (N_0M_1-N_3M_2)}{\sqrt{M_1^2+M_2^2}},\nonumber \\
f'&=&-\frac{2\eta(S_0M_1+S_3M_2)}{\sqrt{M_1^2+M_2^2}},
\end{eqnarray}
for $\eta=\pm1$, 
and various quantities are given in \eqref{M_def} and \eqref{N_S_def}. As in the previous case, we also have two algebraic constraints  
\begin{eqnarray}    
M_1N_3&=&-N_0M_2 \nonumber \\
\textrm{and}\qquad e^{-2f}&=&4(S_0^2+S_3^2)-4(G_0S_0+G_3S_3)^2.
\end{eqnarray}
There are more equations to be solved and also more boundary conditions for scalars to be fine-tuned. After an intensive numerical search, a representative example of possible regular Janus solutions is given in figure \ref{fig5} for $g=3m$ and $m=\frac{1}{2}$. As in the previous case, the solution gives a holographic description of a conformal interface within $N=2$ SCFT in the presence of sources and position-dependent expectation values for both relevant and irrelevant operators. It could be interesting to find field theory descriptions for conformal interfaces dual to all Janus solutions considered in this section. 

\section{Conclusions}\label{conclusion}
We have constructed $F(4)$ gauged supergravity in six dimensions with $ISO(3)\times U(1)$ gauge group by coupling four vector multiplets to the supergravity multiplet. This gauged supergravity can be embedded in type IIB string theory by a consistent truncation on $S^2\times \Sigma$ as shown in \cite{Henning_Malek_AdS7_6}. It turns out that there is only one supersymmetric $AdS_6$ vacuum with $SO(3)$ symmetry which is expected to be dual to an $N=2$ SCFT in five dimensions. We have also given the complete scalar mass spectrum at this $AdS_6$ critical point which encodes information about conformal dimensions of the corresponding operators dual to the scalars. By truncating to $SO(3)\times U(1)\subset ISO(3)\times U(1)$ and $SO(2)\times U(1)\subset SO(3)\times U(1)$ invariant sectors, we have studied a number of holographic RG flows from the conformal fixed point dual to the $AdS_6$ vacuum to different non-conformal phases of the $N=2$ SCFT. Many solutions can be found analytically, but some of them can only be found numerically. Solutions within a subtruncation to three vector multiplets leading to $ISO(3)$ gauge group have also been considered. As usual in this type of solutions, there are singularities at a finite value of the radial coordinate. We have shown that some of these singularities are physically acceptable using the criterions given in \cite{Gubser_singularity} and \cite{maldacena_nogo}.     
\\
\indent In addition, we have also studied a number of supersymmetric Janus solutions by using a curved domain wall ansatz with $AdS_5$ slices. We have considered the solutions within the $SO(2)\times U(1)$ truncation and possible consistent subtruncations. A number of regular Janus solutions interpolating between $AdS_6$ vacua is given numerically. All of these solutions should holographically describe conformal interfaces within the dual five-dimensional $N=2$ SCFT in the presence of sources and position-dependent expectation values for relevant and irrelevant operators dual to different scalar fieds in the bulk matter-coupled $F(4)$ gauged supergravity.  
\\
\indent It would be interesting to identify the dual $N=2$ SCFT in five dimensions and find field theory descriptions of the holographic RG flows given in this paper. Moreover, constructing conformal interfaces dual to the Janus solutions presented here in the framework of the dual five-dimensional $N=2$ SCFT is also worth considering. Working out the complete truncation ansatz of type IIB theory to both $ISO(3)$ and $ISO(3)\times U(1)$ matter-coupled $F(4)$ gauged supergravities is not only interesting in its own right but also useful in uplifting all the solutions found in this paper to ten dimensions. This could lead to possible brane configurations described at low energy by six-dimensional solutions and could also give rise to the precise identification of the dual five-dimensional $N=2$ SCFT. However, as pointed out in \cite{Henning_Malek_AdS7_6}, it is not clear whether there are globally regular solutions in type IIB theory to which these solutions uplift. Finding other types of holographic solutions within the $ISO(3)\times U(1)$ $F(4)$ gauged supergravity is also of particular interest \cite{ISO3_defect} and \cite{ISO3_BH}. Finally, constructing a Euclidean version of Janus solutions given here and using it to holographically study mass deformations of the five-dimensional SCFT along the line of \cite{mass_deform_5D_SCFT} are also interesting.       
\vspace{0.5cm}\\
{\large{\textbf{Acknowledgement}}} \\
This work is funded by National Research Council of Thailand (NRCT) and Chulalongkorn University under grant N42A650263. 


\begin{thebibliography}{99}
\bibitem{Nahm_res} W. Nahm, ``Supersymmetries and their representations", Nucl. Phys. B \textbf{135} (1978) 149-166.
\bibitem{Seiberg_5Dfield} N. Seiberg, ``Five dimensional SUSY field theories, non-trivial fixed points and string dynamics'', Phys. Lett. \textbf{B388} (1996) 753-760, arXiv: hep-th/9608111.
\bibitem{Seiberg_5Dfield3} D. R. Morrison and N. Seiberg, ``Extremal transitions and five-dimensional supersymmetric field theories'', Nucl. Phys. \textbf{B483} (1997) 229, arXiv: hep-th/9609070.
\bibitem{Seiberg_5Dfield2} K. Intriligator, D. R. Morrison and N. Seiberg, ``Five dimensional supersymmetric gauge theories and degenerations of Calabi-Yau spaces'', Nucl. Phys. \textbf{B497}
(1997) 56, arXiv: hep-th/9702198.
\bibitem{maldacena} J. M. Maldacena, ``The large $N$ limit of superconformal field theories and supergravity'', Adv. Theor. Math. Phys. \textbf{2} (1998) 231-252, arXiv: hep-th/9711200.
\bibitem{Gubser_AdS_CFT} S. S. Gubser, I. R. Klebanov and A. M. Polyakov, ``'', Phys. Lett. \textbf{B428} (1998) 105-114, arXiv: hep-th/9802.109.
\bibitem{Witten_AdS_CFT} E. Witten, ``Anti De Sitter Space and holography'', Adv. Theor. Math. Phys. \textbf{2} (1998) 253-291, arXiv: 9802150.
\bibitem{ferrara_AdS6} S. Ferrara, A. Kehagias, H. Partouche, A. Zaffaroni, ``AdS$_6$ interpretation of 5d superconformal field theories'', Phys. Lett. \textbf{B431} (1998) 57-62, arXiv: hep-th/9804006.
\bibitem{D4D8} A. Brandhuber and Y. Oz, ``The D4-D8 brane system and five dimensional fixed
points'', Phys. Lett. \textbf{B460} (1999) 307-312, arXiv: hep-th/9905148.
\bibitem{5DSymmetry_enhanced} D. Bashkirov, ``A comment on the enhancement of global symmetries in superconformal $SU(2)$ gauge theories
in 5D'', arXiv: 1211.4886.
\bibitem{Bergman} O. Bergman and D. Rodriguez-Gomez, ``5d quivers and their AdS(6) duals'' JHEP 07 (2012) \textbf{171}, arXiv: 1206.3503.
\bibitem{Bergman2} O. Bergman and D. Rodriguez-Gomez, ``Probing the Higgs branch of 5D fixed point theories with dual giant gravitons in AdS(6)'', JHEP 12 (2012) \textbf{047}, arXiv: 1210.0589.
\bibitem{Bergman3} O. Bergman, D. Rodriguez-Gomez and G. Zafrir, ``5D superconformal indices at large $N$ and holography'', JHEP 08 (2013) \textbf{081}, arXiv: 1305.6870.
\bibitem{AdS6_CFT5_1} O. Bergman, D. Rodriguez-Gomez and C. F. Uhlemann, ``Testing AdS$_6$/CFT$_{5}$ in Type IIB with stringy operators'', JHEP 08 (2018) \textbf{127}, arXiv: 1806.07898.
\bibitem{AdS6_CFT5_2} M. Fluder and C. F. Uhlemann, ``Precision Test of AdS$_6$/CFT$_{5}$ in Type IIB String  Theory'', Phys. Rev. Lett. \textbf{121} (2018) 17, 171603, arXiv: 1806.08374.
\bibitem{AdS6_CFT5_3} F. Apruzzi, L. Lin and C. Mayrhofer, ``Phases of 5d SCFTs from M-/F-theory on Non-Flat Fibrations'', JHEP 05 (2019) \textbf{187}, arXiv: 1811.12400.
\bibitem{AdS6_CFT5_4} C. F. Uhlemann, ``Exact results for 5d SCFTs of long quiver type'', JHEP 11 (2019) \textbf{072}, arXiv: 1909.01369.
\bibitem{AdS6_CFT5_5} P. B. Genolini, M. Honda, H. Kim, D. Tong and C. Vafa, ``Evidence for a Non-Supersymmetric 5d CFT from Deformations of 5d $SU(2)$ SYM'', JHEP 05 (2020) \textbf{058}, arXiv: 2001.00023.
\bibitem{F4_Romans} L. J. Romans, ``The $F(4)$ gauged supergravity in six-dimensions'', Nucl. Phys \textbf{B269} (1986) 691.
\bibitem{F4SUGRA1} R. D' Auria, S. Ferrara and S. Vaula, ``Matter coupled $F(4)$ supergravity and the AdS$_6$/CFT$_5$ correspondence'', JHEP 10 (2000) \textbf{013}, arXiv:
hep-th/0006107.
\bibitem{F4SUGRA2} L. Andrianopoli, R. D' Auria and S. Vaula, ``Matter coupled $F(4)$ gauged supergravity Lagrangian'', JHEP 05 (2001) \textbf{065}, arXiv: hep-th/0104155.
\bibitem{F4_nunezAdS6} U. Gursoy, C. Nunez and M. Schvellinger, ``RG flows from Spin(7), CY 4-fold
and HK manifolds to AdS, Penrose limits and pp waves'', JHEP 06
(2002) \textbf{015}, arXiv: hep-th/0203124.
\bibitem{F4_flow} P. Karndumri, ``Holographic RG flows in six dimensional F(4) gauged
supergravity'', JHEP 01 (2013) \textbf{134}, Erratum-ibid. JHEP 06 (2015) \textbf{165}, arXiv: 1210.8064.
\bibitem{5DSYM_from_F4} P. Karndumri, ``Gravity duals of 5D N=2 SYM from F(4) gauged supergravity'',
Phys. Rev. \textbf{D90} (2014) 086009, arXiv: 1403.1150.
\bibitem{6D_twist} P. Karndumri, ``Twisted compactification of $N = 2$ 5D SCFTs to three and two dimensions from $F(4)$ gauged supergravity'', JHEP 09 (2015) \textbf{034}, arXiv: 1507.01515.
\bibitem{6D_Janus} M. Gutperle, J. Kaidi and H. Raj, ``Janus solutions in six-dimensional gauged supergravity'', JHEP 12 (2017) \textbf{018}, arXiv: 1709.09204.
\bibitem{6D_Janus_RG} P. Karndumri, ``Janus and RG-flow interfaces from matter-coupled $F(4)$ gauged supergravity'', arXiv: 2405.17169.
\bibitem{AdS6_BH_Minwoo1} M. Suh, ``Supersymmetric $AdS_6$ black holes from $F(4)$ gauged supergravity'', JHEP 01 (2019) \textbf{035}, arXiv: 1809.03517.
\bibitem{AdS6_BH_Zaffaroni} S. M. Hosseini, K. Hristov, A. Passias, A. Zaffaroni,  ``6D attractors and black hole microstates'', JHEP 12 (2018) \textbf{001}, arXiv: 1809.10685.
\bibitem{AdS6_BH_Minwoo} M. Suh, ``Supersymmetric $AdS_6$ black holes from matter coupled $F(4)$ gauged supergravity'', JHEP 02 (2019) \textbf{108}, arXiv: 1810.00675.
\bibitem{AdS6_BH} P. Karndumri, ``New supersymmetric $AdS_6$ black holes from matter-coupled $F(4)$ gauged supergravity'', Eur. Phys. J. Plus 139 (2024) \textbf{858}, arXiv: 2403.01746.
\bibitem{AdS6_defect} P. Karndumri, ``Line and surface defects in 5D $N=2$ SCFT from matter-coupled $F(4)$ gauged supergravity'', arXiv: 2406.18946.
\bibitem{Massive_IIA_onS4} M. Cvetic, H. Lu and C. N. Pope, ``Gauged six-dimensional supergravity from
massive type IIA'', Phys. Rev. Lett. \textbf{83} (1999) 5226-5229,
arXiv: hep-th/9906221.
\bibitem{AdS6_Jan} P. Karndumri and J. Louis, ``Supersymmetric $AdS_6$ vacua in six-dimensional $N=(1,1)$ gauged supergravity'', JHEP \textbf{01} (2017) 069, arXv: 1612.00301.
\bibitem{Henning_Malek_AdS7_6} E. Malek, H. Samtleben and V. V. Camell, ``Supersymmetric $AdS_7$ and $AdS_6$ vacua and their consistent truncations with vector multiplets'', JHEP 04 (2019) \textbf{088}, arXiv: 1901.11039.
\bibitem{AdS6_IIB1} E. D’Hoker, M. Gutperle, A. Karch, and C. F. Uhlemann, ``Warped $AdS_6\times S^2$ in Type IIB supergravity I: Local solutions'', JHEP 08 (2016) \textbf{046}, arXiv: 1606.01254.
\bibitem{AdS6_IIB2} E. D’Hoker, M. Gutperle, and C. F. Uhlemann, ``Warped $AdS_6\times S^2$ in Type IIB supergravity II: Global solutions and five-brane webs'', JHEP 05 (2017) \textbf{131}, arXiv: 1703.08186.
\bibitem{AdS6_IIB3} E. D’Hoker, M. Gutperle, and C. F. Uhlemann, ``Warped $AdS_6\times S^2$ in Type IIB supergravity III: Global solutions with seven-branes'', JHEP 11 (2017) \textbf{200}, arXiv: 1706.00433.
\bibitem{maldacena_nogo} J. Maldacena and C. Nunez, ``Supergravity description of field theories on
curved manifolds and a no go theorem'', Int. J. Mod. Phys. \textbf{A16} (2001) 822,
arXiv: hep-th/0007018.
\bibitem{Gubser_singularity} S. S. Gubser, ``Curvature singularities: the good, the bad and the naked'', Adv. Theor. Math. Phys. \textbf{4} (2000) 679-745.
\bibitem{ISO3_defect} P. Karndumri, ``Twisted compactifications and conformal defects from ISO(3)$\times$U(1) $F(4)$ gauged supergravity'', arXiv: 2410.04403.
\bibitem{ISO3_BH} P. Karndumri, ``Supersymmetric $AdS_6$ black holes from ISO(3)$\times$U(1) $F(4)$ gauged supergravity'', arXiv: 2410.07837.
\bibitem{mass_deform_5D_SCFT} M. Gutperle, J. Kaidi and H. Raj, ``Mass deformations of 5d SCFTs via holography'', JHEP 02 (2018) \textbf{165}, arXiv: 1801.00730.
\end{thebibliography}
\end{document}